\documentclass{jpp}
\usepackage{graphicx}
\usepackage{amsmath}
\usepackage{amssymb}
\usepackage{bm}
\usepackage{color}
\usepackage{accents}
\usepackage{physics}
\usepackage{subfigure}
\usepackage[bookmarks=false, colorlinks]{hyperref}
\usepackage{natbib}
\newcommand\remark[1]{{\color{red}[{\it{#1}}]}}
\usepackage{yfonts}
\newcommand{\beq}{\begin{equation}}
\newcommand{\eeq}{\end{equation}}
\newcommand{\utilde}{\underaccent \sim}

\newcommand{\lra}[1]{\left\langle{#1}\right\rangle}
\newcommand{\del}{\bm{\nabla}}

\newcommand\bmB{\bm{B}}
\newcommand\bmb{\bm{b}}
\newcommand\bmU{\bm{U}}
\def\bmu{\bm{u}}
\newcommand\bmA{\bm{A}}
\newcommand\bma{\bm{a}}
\newcommand\bmx{\bm{x}}
\newcommand\bmK{\bm{K}}
\newcommand\bmk{\bm{k}}

\newcommand{\Gl}{G_l}                % kernel of average
\newcommand{\uGl}{{\utilde G}_l}   

\def\bar#1{\overline{#1}}    % Gaussian averaged mean quantities
\newcommand{\hg}{\hat\gamma}         % gamma operator
\newcommand{\hr}{\hat{\bm r}}   % r hat
\newcommand{\hp}{\hat{\bm \phi}} % phi hat
\newcommand{\sinc}{\text{sinc}} % sinc functoin
\newcommand{\ep}{\epsilon}    % epsilon
\newcommand{\emf}{\bm{\mathcal{E}}} %EMF
\newcommand{\mathO}{\mathcal{O}} %mathcal O
\newcommand{\sgma}{\sigma_{\alpha_{\text k}}}
\newcommand{\sgmb}{\sigma_\beta}
\newcommand{\sgmab}{\sigma_{{\alpha_{\text k}}\beta}}
\newcommand{\alphak}{\alpha_{\text k}} %alpha_k
\newcommand{\alpham}{\alpha_{\text m}} %alpha_m
\newcommand{\errorint}{\sigma^2_{\text{IE}}} %intrinsic error
\newcommand{\errorintb}{\sigma^2_{\text{IE},\overline{B}_i}} %intrinsic error of Bbar
\newcommand{\errordf}{\sigma^2_{\text{FE}}} %filtering error
\newcommand{\lopt}{l_{\text {opt}}} %l_optimal
\newcommand{\lsm}{l_{\text{s}}} %l_s
\newcommand{\llg}{l_{\text{L}}} %l_L
\newcommand{\ksm}{k_{\text{s}}}
\newcommand{\klg}{k_{\text{L}}}
\newcommand{\rsun}{r_\odot}

\newcommand{\Ra}{R_\alpha}
\newcommand{\Rw}{R_\omega}
\newcommand{\RU}{R_U}

\newcommand{\sigs}{\sigma_{\text{s}}}
\newcommand{\sigL}{\sigma_{\text{L}}}
\newcommand{\Pe}{P_\text{e}}
\newcommand{\Pa}{P_\text{a}}
\shortauthor{H. Zhou, E.G. Blackman, L. Chamandy}
\title{Derivation and precision of mean field electrodynamics with mesoscale fluctuations}
\shorttitle{Derivation and Precision of Mean field Electrodynamics}

\author[H. Zhou, E.G. Blackman, L. Chamandy]
{H\ls O\ls N\ls G\ls Z\ls  H\ls E\ \ls  Z\ls H\ls O\ls U$^{1,2}$
 \thanks{Email address for correspondence: hzhou21@ur.rochester.edu}\ns
 E\ls R\ls I\ls C\ls \ G.\ \ls B\ls L\ls A\ls C\ls K\ls M\ls A\ls N$^{1,2}$
\thanks{Email address for correspondence: blackman@pas.rochester.edu}\ns
L\ls U\ls K\ls E\ls  \ C\ls H\ls A\ls M\ls A\ls N\ls D\ls Y$^{1}$
\thanks{Email address for correspondence: lchamandy@pas.rochester.edu}}
\affiliation{$^1$ Department of Physics and Astronomy, University of Rochester,
Rochester, NY, 14627, USA\\
[\affilskip]
$^2$ Laboratory for Laser Energetics,  University of Rochester, Rochester NY, 14623, USA\\[\affilskip]
}
\pubyear{}
\volume{}
\pagerange{}
\date{?; revised ?; accepted ?. - To be entered by editorial office}

\begin{document}

\maketitle

\begin{abstract}

Mean field electrodynamics (MFE)  facilitates  practical modeling of secular, large scale properties 
of astrophysical or laboratory 
%of magnetohydrodynamic or plasma 
systems with fluctuations.
Practitioners commonly assume   wide scale separation between mean  and  fluctuating quantities,   to justify equality of ensemble and spatial or temporal averages.
Often however,   real systems  do not exhibit 
such     scale separation. This raises  two  questions: ({\bf I}) what are the appropriate generalized equations of MFE in the presence of  mesoscale fluctuations? ({\bf II}) how precise are  theoretical predictions from  MFE?  
We address both  by first  deriving the equations of MFE for different  types of averaging,  along with  mesoscale correction terms that depend on the ratio of averaging scale to variation scale of the mean. 
We then show that even if these terms are small,  predictions of MFE 
can  still have a significant  precision error. 
This error has an {\it intrinsic}  contribution from the dynamo input parameters   and a {\it filtering} contribution   from  differences in the way  observations and theory are projected  through the 
 measurement kernel.
%difference  between the  observational and theoretical 
Minimizing the sum of these  contributions can produce an optimal
scale of averaging that makes the theory maximally precise.
The precision error  is important to quantify when comparing to observations  because it quantifies
the resolution of  predictive power.
  %The errors depend on the 
  %on the aforementioned scale ratio,  the strength of the fluctuations, and their spectra. 
   %MFE offers  less precision than  the averaging over particles that produces the starting MHD equations because  the pertinent  ensembles   are much smaller.  
 We exemplify these principles for galactic dynamos, comment on broader implications, and identify
 possibilities for further work.

\end{abstract}

\begin{PACS}
%Authors should not enter PACS codes directly on the manuscript, as these must be chosen during the online submission process and will then be added during the typesetting process (see http://www.aip.org/pacs/ for the full list of PACS codes)
\end{PACS}

%\remark{EGB testing new   command which may be useful for our comments instead of\%} 

%\blue{EGB testing new color commands which may be useful}

%\green{EGB testing new color commands which may be useful}

%=======================================================

\section{Introduction}\label{sec_intro}
Mean field electrodynamics (MFE) is a powerful tool for semi-analytical modeling of  large scale 
or secular behavior of magnetic fields and  flows in 
 magnetohydrodynamic and plasma systems
with spatial or temporal disorder 
\citep[e.g.][]{RobertsSoward1975, KrauseRadler1980, RSS1988, BrandenburgSubramanian2005, KleeorinRogachevskii2008, KleeorinRogachevskii2009, Blackman2015}.
As its name indicates,  in MFE physical variables such as the magnetic field $\bmB$ and velocity  $\bmU$ are  decomposed into mean  and fluctuating parts and the equations for the means are derived. The ubiquity of turbulence in astrophysics renders 
MFE essential for practical comparison between theory and observation.
Mean field magnetic dynamo theory is a  prominent example  
of MFE.  Standard axisymmetric accretion disk theory with `turbulent'  transport is  
another example, although many practitioners  use the theory without
recognizing that it is only valid  as a mean field theory, and in fact one that should be
coupled to mean field dynamo theory \citep{BlackmanNauman2015}.
%of MFE: turbulence is only consistent with axisymmetry  after averaging over fluctuations. That standard accretion theory is itself a mean field theory is less  widely appreciated by practical users than for the case of mean field dynamo theory. 
%For disks, dynamo theory and standard accretion theory  are  in fact artificially decoupled components of what should be a unified theory.
By itself, the term MFE does not specify a single  set of approximations or  method of averaging. If a system shows large scale  field or flow patterns the question is not whether MFE is correct but what is the  most 
%HZ30
%correct 
appropriate MFE.

Specific averaging methods   include the ensemble average (over a very large number of accessible microstates),
spatial averages (like box or planar averages), and  time averages. 
%to name a few.  
Calculations are usually simplified by utilization of  Reynolds rules, namely,
the linearity of averaging,
the interchangeability of differential and average operations,
%\remark{HZ edited below}
and that averaged quantities behave like constants in averages
(e.g., an averaged quantity is invariant if averaged more than once,
and the average of the product of a quantity and a mean quantity is equal to the product of the mean of these two quantities.)
The ensemble average respects the full Reynolds rules, and  is  commonly favored 
\citep[e.g.][]{RobertsSoward1975, BrandenburgSubramanian2005}.    
%\subsubsection {\it Ensemble average}. 
%	The most widely used average is ensemble average, a concept borrowed from statistical mechanics. 
In the ensemble average, means are obtained by averaging over an ensemble consisting of a large number of identical systems prepared with different initial states. 
	Fluctuations then have zero means by definition, and statistical properties of all mean physical quantities, such as the turbulent electromotive force (EMF), are determined once the partition function is known. 
	There might seem to be no need to invoke the assumption of large scale separation,  but   the detailed statistical mechanics and partition function are rarely  discussed	in the MFE
	%	\remark{HZ added references in footnotes here}
	context\footnote{
		For hydrodynamic ensembles, see \cite{Kraichnan1973};
		\cite{Frisch1975} has studied MHD ensembles at absolute statistical equilibrium;
		more applications of ensembles in MHD systems can be found in \cite{Shebalin2013} and the references therein.
},
%	\remark{HZ added below, EB tweaked}
	%and due to the same reason,
so it is unclear how to calculate variations of these systems from first principles.
	
	%We therefore conclude that ensemble average and a large scale separation are equally of the essence in the traditional MFE. 	

%A discrepancy then comes in between theoretical predictions and our data in hand: 
%The former uses ensemble average while the latter ones do not. 
   
%Replicas of the same event are unavailable in astrophysical observations, since even the smallest dynamical time scale, the turbulent turnover time, still exceeds the age of modern astrophysical researches; 
%thousands or millions times of experiments or simulations are also very much time-consuming or even impossible.
Correlation functions in magnetohydrodynamics (MHD) are usually computed  from the equations of motion, either  in configuration space or Fourier space 
\citep[e.g.][]{PouquetFrischLeorat1976, RSS1988, BlackmanField2002}.
Spatial or temporal averages are  the most directly relevant choices when analyzing simulations, laboratory experiments, or astrophysical observations. 
These averages can however, explicitly break the Reynolds rules
in the absence of large scale separation.
For example, a planar average in the horizontal plane will destroy any spatial dependence in, say, the $x-y$ plane and leave physical quantities solely a function of $z$, therefore variant when interchanged with $\partial_x$ or $\partial_y$ unless the boundary conditions are periodic.
Another example is a weighted-averaging over a local small volume,  %as to be discussed in the current paper, 
which  retains full coordinate dependence at the price of a double-averaged quantity which is generally unequal to its single-averaged value as we will later discuss in detail.
To avoid these complications,  MFE practitioners  typically assume 
 that the system to which the theory is being compared has a large  scale separation between fluctuating and mean quantities. The Reynolds 
rules are then quasi-justified  for  spatial and temporal averages, and are deemed 
to be good  approximations to the ensemble average  \citep{BrandenburgSubramanian2005}. 
% the MFE theory constructed using ensemble averages to experimental or observational data. 

%\remark{EGB: The sentence below has been moved here.}
Some effects of turbulence on astrophysical observables have been  discussed \citep{Burn1966,Spangler1982, Eilek1989a, Eilek1989b, Tribble1991, Sokoloff1998}, but 
 the mean or ordered fields were typically defined  explicitly or implicitly via ensemble averages.
 Here we focus on the problem  that  real systems 
%such as stars, galaxies, and accretion disks,
do not typically have a large  scale separation between fluctuations and large scale quantities, and thus equating ensemble and spatial averages above can be questioned.
If $\lsm$ and $\llg$ are the characteristic  lengths of small- and large-scale fields, 
galaxies, for example, may have  
%choose $l$  such that $\lsm<l<\llg$
$\llg\simeq 1\text{\ kpc}$ 
and  $0.05 \text{\ kpc}\le \lsm \le 0.1\text{\ kpc}$ so that $\lsm/\llg \ge 1/20$ which is not infinitesimal.
%\remark{HZ added the sentence below as suggested by referee 1}
As another example, in  one of their solar dynamo models, \cite{Moss2008} have introduced a dynamo 
%alpha
coefficient with long-term variations and a correlation time
%\remark{HZ added ``of turbulent fields"}
of turbulent fields set to be of the same order as the period of the solar magnetic activity.
In this case, the ratio of the mean to fluctuating time scales would be $\sim1$. 
 %EB  not $\sim 10^9\text{\ s}$ but $10^7$ yr for these structures.
% The approximate equivalence between ensemble and other kinds of averages needs to be 
%investigated
%\remark{HZ added below}
Finite scale separation in time scales is equivalent to $\lra{\bmB}\neq \int_t^{t+T} dt\ \bmB(t)$ where $T$ is a time scale much greater then the eddy turnover time, but still much smaller than the time scale of mean fields.
This implies the system is non-ergodic.
For more detailed discussions about non-ergodicity of MHD systems, see \cite{Shebalin1989, Shebalin2010, Shebalin2013a} and the references therein.
We are  thus led to two specific questions: {\bf (I)} in  the presence of  intermediate or mesoscale fluctuations
what are the correction terms to standard ensemble averaged MFE? and
{\bf (II)} what precision
%\remark{EGB: deleted error}
does this imply when comparing the theory to observations?

%Unfortunately, the ensemble-averaged and fluctuating fields do not necessarily respect a %large scale separation. 
%It is non-trivial to rule out the possibility that an ensemble-averaged field has small %scale structures, or its fluctuation evolves over large scales.
%On the other hand, in both simulations and observations, we do observe physical quantities %of different scales and a clear separation of slowly-evolving large-scale fields and %fast-evolving small-scale fields. 
%To what extent, then, does the conventional MFE match the observations, if the relation %between ensemble-averaged fields from the theory and observed large-scale fields is unclear?

To address question ({\bf I}),  we compare the standard MFE equations from ensemble averaging to those formally derived using a spatially local average 
%(hereafter called `local average') 
when
the scale of averaging is not arbitrarily smaller than the mean field gradient scales.
We define  spatial averages as  convolutions between the total field and a  kernel with a prescribed scale of averaging $l$ such that  $\lsm<l<\llg$  \citep{Germano1992}.
%HZ4
%EB10 looks good minor tweaks
Such  `coarse-graining' techniques have been applied to  hydrodynamic turbulence \citep{Leonard1974, Meneveau2000,EyinkAluie2009}, as well as MHD turbulence \citep{AluieEyink2010, Aluie2017}.
\cite{Gent2013}  used a Gaussian kernel for averaging in simulations to explore  scale separation of magnetic fields.  \cite{Frick2001}  used a  mathematically similar method, wavelet transforms, for the analysis of galactic images.  Relevant kernels are localized in both configuration and Fourier space to  filter out small scales. Here we go beyond previous work and  derive corrections to   standard MFE which depend on the ratio $(l/\llg)^c$, 
%If $\lsm$ and $\llg$ are the characteristic  lengths of small- and large-scale fields, respectively, we 
%choose $l$  such that $\lsm<l<\llg$ to include a number eddies to give a statistical description of turbulent quantities, and still be able to figure out large-scale configurations of  mean fields. 
where  the power   $c$ depends on the   choice of 
 kernel.  For  $l/\llg\ll1$ the standard MFE equations are recovered.

%EB3 added below
Another way to describe the importance of mesoscale  fluctuations for  MFE is that 
contributions to averages are non-local, requiring weighing over a kernel
of  finite spatial or temporal range.  In this respect, what we do here  differs from 
 \cite{Rheinhardt2012}, even though they  also  motivate their work by recognizing a  need to account for non-locality.  Their focus is on empirically extracting  from simulations
the kernel of proportionality relating the turbulent EMF and the mean magnetic field, and constraining  an  ansatz for that kernel when the mean magnetic field is defined with a planar average.  In contrast,   we   derive  corrections that directly arise from the mean field averaging procedure  itself, and identify the  lowest order correction terms resulting from distinct choices of the averaging kernel when Reynolds rules are violated.
As we  discuss later,  the  approach  of \cite{Rheinhardt2012}
 can actually  be viewed as   semi-empirically testing the turbulent closure in MFE.
 %, given some prescribed averaging method for mean field quantities. 
 %It is not aimed at corrections that arise from violation of Reynolds rules.  
 % They
   %Given their free parameters and length scales used, 
   %    have   not  yet demonstrated the necessity of  their MFE equations with two closure parameters
  % beyond  the one parameter minimal tau closure approximation for triple correlations.
      
To address question ({\bf II}) above, the precision of MFE in the presence of mesoscale fluctuations, we  identify two types of errors:
(i) the `intrinsic error' (IE) of the mean fields that arises  from the uncertainties to the input parameters of the mean field equations, and
(ii) the `filtering error' (FE)  that results if the theoretical averaging procedure does not match
that for values extracted from the observational data.
%\remark{EGB: tweaked above}
% total field filtered through the measuring kernel. We call this the "double filtering" error (FE).
 %EB20 change above ok? 
% results for excluding large wavenumber modes from mean fields?
%\remark{HZ added below}
As we will see in Sec. \ref{sec_twoerrors}, when using ensemble averages, the IE vanishes and the FE is finite but unquantifiable if partition functions are unknown.
 For the IE in our formalism, 
%type (i),  
we identify the importance of the ratio 
 $l/\lsm$, where $\lsm$ is the  integral  (energy dominating)  scale of the
% \remark{HZ changed below}
 turbulent magnetic field.
% \remark{EGB:  what if magnetic and kinetic energy dominating scales are different?}
% Note that $\lsm$ depends on how large and small scales are defined  and thus is a function of $l$.
%\remark{EGB removed above sentence}
%EB20 tweaked above
This ratio emerges   because 
% for the parameters in mean field equations, 
contributions to the error about the mean from fluctuations vary  as $\sim N^{-1/2}$ where $N\simeq (l/\lsm)^3$ is the number of eddies contained in an averaged cell. %For fixed  $\lsm$, a large $l$ means $N^{-1/2}\rightarrow 0$ and the error on the mean values adopted in equations would be small. 
%As in the previous paragraph however, the story changes for more modest  scale separations.
For the FE,   the ratio $L/\lsm$ is most important, where  $L$ is the scale of average associated with the observation method, and in general differs from  $l$.
%In short, the smaller the averaging scale $l$, the smaller the $N$ and the larger the error.
Although $\lsm$ increases  with increasing $l$ because $\lsm$ is roughly the average scale of modes with wavenumbers $\leq2\pi/l$, the dependence is  weak if the small scale turbulent 
spectrum of the magnetic field peaks near $\lsm$.
As a result, the ratio $l/\lsm$ is roughly proportional to $l$  whereas $L/\lsm$ decreases as $l$ increases. That the IE and FE  have  complementary dependences on $l$  implies that their
sum may  have an optimal scale of averaging that minimizes the total error.
We will show that both types of precision errors  are quantifiable, and can be significant in galaxies for example.  

In section \ref{sec_average} we introduce the local spatial averages using kernels, and formally derive correction terms when  the Reynolds rules are not exactly obeyed.
In section \ref{sec_dynamo} we apply these results to derive the generalized dynamo equations of MFE
and 
%by re-formulating  MFE equations with corrections based on kernels, and the 
show that the mesoscale correction terms are  in fact generally small using order-of-magnitude estimates. 
%EB3 added below
We also contrast our method and compare our equations  to the  dynamo equations of \cite{Rheinhardt2012}.
In section \ref{sec_twoerrors} we present a general discussion on the two types of uncertainties aforementioned.
%\remark{EGB: tweaked below}
In 
section \ref{sec_precision} 
%we derive more specific 
%orms of precision  of the  MFE and find  that the error
%can be quite sizeable.
we show how to compute the total error in the specific case of comparing MFE to Faraday rotation (FR)  measurements and  apply this to different galactic viewing angles in section \ref{sec_application}.
We conclude in section \ref{sec_conclusion}.

\section{Averaging in MFE using kernels}\label{sec_average}
In this section, we introduce the general  formalism for averaging using kernels, preparing for the reformulation of MFE in the next section. 

%EB minor notational stuff:  (1)  i now  distinguish configuration space quantities from their transforms using tildes underneath the latter  throughout, I defined the command \utilde to help. (2)  We might later consider changing  the notation of $A$ and $B$ for the vectors to something like Q and S throughout  wont immediately make people think of vector potential and magnetic field  (as an aside, only the former is a true vector :))

\subsection{General formalism}
\label{sec:2.1}
As per  standard  MFE practice, we separate any vector field $\bmA$ into a mean part $\bar\bmA$ and a fluctuation part $\bma$,
\beq
\bmA(\bmx)=\bar\bmA(\bmx)+\bma(\bmx).
\eeq
The mean part is defined via
\beq
\bar\bmA(\bmx)=\Gl(\bmx)*\bmA(\bmx)=\int d^3 x'\ \Gl(\bmx-\bmx')\bmA(\bmx'),
\label{mean}
\eeq
where `$*$' denotes a convolution. 
The filtering kernel  $\Gl(\bmx)$ is a prescribed function with a characteristic  scale of averaging $l$, satisfying $\lsm< l_{\text{eff}}(l) < \llg$,
where $l_{\text{eff}} $ can be viewed as the configuration space dividing scale between large and small-scale fields.
We define $l$ such that 
 $l=l_{\text{eff}}$  
 % theoretical scale
  for our analytic derivations.\footnote{
	The choice of amplitude in the filter function that separates mean from fluctuations and   defining the relation between  $l$ and $l_{\text{eff}}$ is not unique.
	For a Gaussian average,  taking $\utilde G(\bmk)=e^{-k^2 l^2/8\pi^2}=1/2$ as  the dividing line implies that $l_{\text{eff}}=l/\sqrt{2\ln 2}$ separates large- and small-scale fields  in configuration space \citep[as in][]{Gent2013}.
	If instead we use $\utilde G(\bmk)=1/e$,  then $l_{\text{eff}}=l/\sqrt{2}$.
	We  adopt $l_{\text{eff}}=l$, but note  that
 different criteria for the dividing line can lead to a constant multiplicative factor on  $l$.	   If our averaging scale were based on  a real space choice such as  telescope beam width for $l_{\text{eff}} $, leading us to set the exponent in equation (\ref{gaussianinx}) to say 1/2, then
$l_{\text{eff}}=l\sqrt{\ln 2 / 2}/\pi=1/2$.  Beam width may not  however, determine the most appropriate theoretical choice of $l_{\text{eff}}$  for a given magnetic energy spectrum.
%However would  have to be checked that this consistent with a wavenumber that lies in  the spectralvalley.},
}
%\remark{EGB: edited above inequality and sentence and the footnote to accomodate luke comment}
%HZ30
%EB30 good..tweaked slightly
%There may cirmst for example, the beam width of a physical telescope.
We have assumed that the  system under consideration is  statistically homogeneous and isotropic on scales  $\le l$,  
so that $l$ is independent of location and $\Gl(\bmx)$ is isotropic. 
For  anisotropic or inhomogeneous systems, $\Gl(\bmx)$ could  be anisotropic and $l$ could be a function of spatial coordinates.

%HZ0907
We use the following definition of the Fourier transform:
\beq
\mathcal{F}[f(\bmx)](\bmk)
=\utilde f(\bmk)
=\int d^3x\ f(\bmx)e^{-i\bmk\cdot\bmx},
\label{eqn_FT}
\eeq
and
\beq
\mathcal{F}^{-1}[\utilde f(\bmk)](\bmx)
=f(\bmx)
=\frac{1}{(2\pi)^3}\int d^3k\ \utilde f(\bmk) e^{i\bmk\cdot\bmx},
\eeq
and therefore the Fourier transform of $\bar\bmA$ is given by
\beq
\bar{\utilde \bmA}(\bmk)
=\uGl(\bmk)\utilde\bmA(\bmk).
\label{2.3}
\eeq
Unlike  idealized ensemble averages,  equation (\ref{2.3}) implies   $\overline{\bar\bmA}\ne \bar\bmA$ 
since $\uGl^2(\bmk)\neq \uGl(\bmk)$ unless ${\utilde G}(\bmk)=0$ or $1$,
%\remark{EGB edited wording here}
such as for  a step function in Fourier space.
%HZ6
%EB11 is the following tweak ok:
However,
 interchangeability of differential and average  operations, as commonly invoked, is manifest in Fourier space since $k_i [\uGl(\bmk) \utilde A_j(\bmk)]=\uGl(\bmk)[k_i \utilde A_j(\bmk)]$.

The  kernel $\Gl(\bmx)$ must meet several requirements for a practical  
 mean field theory.
First, it should be a spatially local function that decreases rapidly for  $|\bmx|\gtrsim l$,   being that it is used to extract a filtered value at a scale $l$. 
%EB do you agree with my revised wording in above sentence.  later we say that  we demanded $\Gl$ to be monotonically decreasing in k, so i added it, 
%HZ yes
Complementarily, its Fourier transform $\uGl(\bmk)$ should also monotonically decrease and vanish for large  $|\bmk|$.
% in order to filter out small-scale fields in configuration space. 
Furthermore, in the limit $l\to 0$, $\uGl(\bmk)$ approaches unity, since no filtering is needed for large scales.
% essentially no average is made then.
Thus, $\uGl(\bmk)$ can be expanded around 
%HZ30
%unity
$|\bmk|=k=0$ when $|\bmk l|/2\pi=|\bmk|/k_l$ is small compared to unity, yielding
\beq
\uGl(\bmk)=1-\utilde\gamma+\mathO(\utilde\gamma^2),
\label{gamma2.4}
\eeq
where $\utilde\gamma$ is a small parameter related to $|\bmk|/k_l$, and the minus sign is for future convenience. 
Note that $\utilde\gamma$ is independent of 
%HZ30
%$\hat \bmk$ 
the direction of $\bmk$ due to isotropy. 

The inverse Fourier transform of $\utilde\gamma$ is an operator $\hg$ which is  determined by
%\remark{HZ}
%specifying boundary conditions of fields. 
\beq
(\hg f)(\bmx)=\mathcal{F}^{-1}[\utilde\gamma(\bmk)\utilde f(\bmk)](\bmx).
\eeq
Hereafter we  assume that these fields  are either vanishing or  
periodic at the spatial boundaries,
%\remark{HZ}
and therefore any $\utilde\gamma(\bmk)$ proportional to a  power of $i\bmk$ 
%\remark{EGB: edited wording just before}
is simply translated to a $\hg$ which is a spatial derivative raised to the corresponding power.
%EB my edit above ok?  but later i think you actually do use "vanishing"  vanishing AND periodic seems restrictive...
%HZ yes I think either vanishing or periodic is correct. I might have a typo..
When applied to a quantity $Q$ with   smallest characteristic scale  $l_\text{ch}>l$, the order-of-magnitude estimate yields
%EB11 added > l above
\beq
\hg Q\sim\left(\frac{l}{l_\text{ch}}\right)^c Q
\label{gamma_hat}
\eeq
with $c$ being a positive number that  depends on the specific choice of kernel.
%EB added last part, above ok?
%HZ yes

\subsection{Expressions for averages of fluctuations and double averages}
Here we obtain  formulae for  averages of fluctuations and
double averages, both of which do not strictly 
%HZ30
%vanish 
obey Reynolds rules in the presence of mesoscale fluctuations. 
%HZ30
In particular, the averages of fluctuations do not vanish and the double averages will not agree with single-averaged values.
%The expressions are general in that they apply  for any   kernel $\Gl$ meeting the requirements described in the previoussubsection. 
%\remark{EGB: Deleted previous sentence in light of the appendix.}
 We will  use the expressions in  subsequent sections.
 
  %EB any, or any meeting the aforementioned requirements..

We first derive an expression for the mean of  fluctuations, namely
 %of $\bmA$, 
 $\bar\bma=\bar\bmA-\bar{\bar\bmA}$. This  vanishes in conventional MFE using the ensemble average, but not    for spatial averages.
In Fourier space, by definition,
\beq
\bar{\utilde\bma}(\bmk)=\uGl(\bmk){\utilde \bmA}(\bmk)-\uGl^2(\bmk){\utilde \bmA}(\bmk)
=(1-\utilde{G}_l)\bar{\utilde\bmA}.
\label{mean_of_fluc}
\eeq
%EB11 the statement below, depends on assumptions about spectrum of $\bar{\utilde \bmA}$
Since $\bar{\utilde \bmA}$ is a large-scale quantity, it decays rapidly when  $|\bmk|/k_l\gg1$. 
Therefore we can expand the RHS of equation (\ref{mean_of_fluc}) for $|\bmk|/k_l\ll 1$ using equation (\ref{gamma2.4}) to obtain
%EB12: change to previous line ok? 
\beq
\bar{\utilde\bma}(\bmk)=\left[\utilde\gamma+\mathO(\utilde\gamma^2)\right]\bar{\utilde\bmA}.
\eeq
In configuration space, this implies
\beq
\bar\bma(\bmx)=\bar\bmA-\bar{\bar\bmA}=
\left[\hg+\mathO(\hg^2)\right]\bar\bmA
\sim\left(\frac{l}{\llg}\right)^c\bar\bmA,
\label{littleave}
\eeq
if  $\bar\bmA$ has a characteristic variation scale of $\llg$.
Equivalently
\beq
\bar{\bar{\bmA}}=\left[1-\hg+\mathO(\hg^2)\right]\bar\bmA.
\label{fml1}
\eeq
To recover conventional MFE, we  simply take the limit $l/\llg\rightarrow0$ and get $\bar\bma=\bm 0$.

Next, we obtain an expression for  the mean of the product of two fields, $\bar{AB}$, in terms of the mean fields.   Here $A$ and $B$ can be either two scalar fields or the components of some vector fields.
%\remark{HZ edited below}
We adopt a two-scale approach, assuming that the fields have double-peaked spectra and scale separations are large but finite, i.e., we relax the assumption of infinite scale separation in conventional approaches (for details see appendix \ref{appx1} where the valid range of scale separation is quantified).
Other closures may include a test filtering process like that used in the Smagorinsky model \citep{Smagorinsky1963, Germano1991, Lilly1992}.
%\remark{EGB3: add refs above}

By straightforward expansion we have
\beq
\bar{AB}=\bar{\bar A\ \bar B}
+\bar{a\bar B}+\bar{\bar A b}
+\bar{a  b}.
\label{exp}
\eeq
We refer  to the terms on the RHS of equation (\ref{exp}) as $T_1$, $T_2$, $T_3$ and $T_4$, respectively. 
%\remark{HZ changed below}
The calculation of $T_1$ involves only mean quantities, but for practical purposes, it is  convenient to make some further approximations to avoid integro-differential equations.
If $\bar A$ and $\bar B$ both have a characteristic 
scale of variation
%\remark{EGB edited to be more precise}
%wavenumbers
 $\llg$, then the spectrum of the Fourier transform of their product will roughly extend to $k=2\klg$.
If the scale of average satisfies $2\klg l/2\pi=\klg l/\pi\ll1$, we can   use equation (\ref{fml1}) to calculate $T_1$, namely, 
\beq
\bar{\bar A\ \bar B}=\left[1-\hg+\mathO(\hg^2)\right](\bar A\ \bar B).
\label{T1}
\eeq

The Fourier transform of $T_2$ is
\beq
{\utilde T}_2(\bmk)
=\uGl(\bmk)\left[{\utilde a}(\bmk)*\bar {\utilde B}(\bmk)\right]
=\uGl\{[(\uGl^{-1}-1)\bar {\utilde A}]*\bar {\utilde B}\},
\label{T2}
\eeq
 where we have used
% \remark{HZ edited below}
 the definition $\utilde\bma=(1-\uGl)\utilde\bmA$.
The convolution of two quantities with characteristic wavenumbers $k_1$ and $k_2$ will yield wavenumbers $k_1\pm k_2$. 
Note that   $\Gl$ is outside of the square brackets 
%EB do we need the caveat above, since $\Gl$ is itself assumed to be small at high wavenumber?
%HZ I think we don't have to expand $(\Gl^{-1}-1)A$ because $\Gl$ naturally filters out large wave numbers. But the statement above frees us from worrying about the large wave number part of $(\Gl^{-1}-1)A$ and allows us to do the expansion which helps simplify calculations.
%EB2 i meant that maybe we dont need to say "With \GL outside of the brackets" changed wording slightily...
%HZ2 Yes I agree
and so with periodic or vanishing boundary conditions, 
%EB so you did mean vanishing periodic earlier, not vanishing or periodic...?
%HZ no it should be an "or". And I'm wondering why $\bar B$ vanishing at large k is determined by choice of B.C.?
%EB2 sorry, i re-edited above
 only the low wavenumber part of  the factor $(G^{-1}-1)\bar A$ survives 
 on the RHS of equation (\ref{T2}).
% \remark{HZ added belew}
 (The validity of this approximation is discussed in more detail in appendix \ref{appx1}.)
 We therefore expand $\Gl^{-1}$ in a Taylor series, yielding
\beq
(\uGl^{-1}-1){\bar{ \utilde A}}=\left[\utilde\gamma+\mathO(\utilde\gamma^2)\right]\bar {\utilde A},
\eeq
which upon Fourier inversion then implies
\beq
T_2=\bar{a\bar B}=\bar{\bar B\left[\hg+\mathO(\hg^2)\right]\bar A}.
\label{fml2'}
\eeq
Similarly, for $T_3$ we obtain
\beq
T_3=\bar{b\bar A}=\bar{\bar A\left[\hg+\mathO(\hg^2)\right]\bar B}.
\label{fml3'}
\eeq
%In appendix \ref{appx1} we discuss the validity of the approximation (\ref{fml2'}) and (\ref{fml3'}) in more detail.
The sum $T_2+T_3$,  using equation (\ref{T1}), is then
%\remark{HZ edited below}
%HZ6 added the gamma' part
%EB11 great
\beq
\bar{a\bar B}+\bar{b\bar A}
=\left[\hg+\mathO(\hg^2)\right]\left(\bar{\bar A\ \bar B}\right)-\bar{\hg'(\bar A,\bar B)}
=\left[\hg+\mathO(\hg^2)\right]\left(\bar A\ \bar B\right)-\hg'(\bar A,\bar B),
\label{T1T2}
\eeq
where $\hg'$, a binary operator, is  introduced  to account for the violation of the distribution rule of $\hg$;
that is,
\beq
\hg'(A,B)
=\hg(AB)-(A\hg B+B \hg A).
\eeq
Note that
%\remark{HZ edited below}
$\hg'(A,B)$ has the same order of magnitude as $B\hg A$ or $A\hg B$ if $A$ and $B$ have the same characteristic length scale.
%\remark{EGB: something missing in previous sentence--do you mean "$B\hg A$ or $A\hg B$"? but that doesn't seem quite what you meant either}

Combining equations (\ref{exp}), (\ref{T1}) and (\ref{T1T2}) we obtain 
\beq
\bar{AB}=[1+\mathO(\hg^2)]\left(\bar A\ \bar B\right)-\hg'(\bar A,\bar B)+\bar{ab}.
\label{fml2}
\eeq
%Note the important result that this is identical to 
% that the sum of the first three terms on the RHS of equation (\ref{exp}) is identical to 
%what  ensemble averages give if we ignore terms of higher than linear in $\hg$.
Furthermore, it can be verified using equation (\ref{fml1}), (\ref{fml2'}) and (\ref{fml2}) together that
\begin{align}
\bar{A\bar B}
=&\bar A\ \bar{\bar B}-\hg'\left(\bar A,\bar{\bar B}\right)+\bar{a\bar b}\notag\\
=&\bar A(1-\hg){\bar B}-\hg'\left(\bar A,(1-\hg){\bar B}\right)+\bar{\bar b\hg\bar A}+\mathO(\hg^2)\notag\\
=&\bar A(1-\hg)\bar B-\hg'\left(\bar A,\bar B\right)+\mathO(\hg^2)\notag\\
=&(1-\hg)\left(\bar A\ \bar B\right)+\bar B\hg\bar A+\mathO(\hg^2).
\label{fml3}
\end{align}

\subsection{Comparison to previous work}
%\remark{HZ added this section, EB tweaked}
%
%Leonard term: (6.5) in \cite{Leonard1974}, (3.16) in \cite{Yeo1987}. Taylor-expand mean fields.
%
%Cross term like (2.22): Sec. 3.3.2 in \cite{Yeo1987}. Express $u'$ in $\bar u$.
%
Expressing a turbulent field as $\utilde a=(1-\uGl)\utilde A$ is equivalent to applying a high-pass filter on $A$, as has been discussed in \cite{Yeo1987}.
  In \cite{Yeo1987} all fields are expressed in terms of mean fields and their derivatives (Yeo-Bedford expansion), including small-scale fields, so the approach facilitates a closure   in their context of the inertial range for Large Eddy Simulations (LES). 
 In  our approach, we focus on the large scale mean fields, not the inertial range.
We keep two-point correlations of turbulent fields [$\bar{ab}$-like terms in equation (\ref{fml2})] but use a separate closure for triple correlations 
  [compare  equations (5.8) to (5.11)  of  \cite{Yeo1987} to our equation (\ref{fml2})].
In our formalism, $\hg$ terms enter as corrections to capture finite-scale separation effects, facilitating   comparisons to conventional approaches (e.g.,  ensemble averages), while allowing different closures.
Also, 
%Two additional differences between our work and that of \cite{Yeo1987}; 
%i) we adopt a two-scale approach and focus on studying large-scale mean fields, whereas  \cite{Yeo1987} applies in the inertial range of turbulent fields; and
 \cite{Yeo1987}  use a Gaussian kernel, whereas our discussions in the previous sections apply to any kernel meeting the requirements in Sec. \ref{sec:2.1}.

%In addition, the motivation and physical interpretation of the present work and \cite{Yeo1987} are different.
%Yeo-Bedford expansion aims at closing the fluid equations, and is convenient in simulations.
%However, using a separate closure (as we will do in Sec. \ref{sec_dynamo}) retains more insight of physics (e.g., the role of helicity).

\subsection{Unifying   different  averaging methods using kernels}
\label{sec:2.3}
Here we discuss commonly used averages and their kernel forms (if possible).
%from which the validity for a mean field theory can be investigated.  
Recall from above that in order to accurately capture large-scale features, a suitable kernel for mean field theories should at least be monotonically decreasing in Fourier space.

\subsubsection{\it Gaussian average} 
	For  isotropic and homogeneous turbulence, the Gaussian kernel is defined as
	%EB deleted "microscopic" here and earlier in the paper, as it is a term which usually applies at the particle level  not fluctuations  of bulk flow.	
	\beq
	\Gl(\bmx)=\left(\frac{k_l^2}{2\pi}\right)^{3/2} e^{-k_l^2 |\bmx|^2/2},
	\label{gaussianinx}
	\eeq
	where $k_l=2\pi/l$.
	It is then evident that $\bar\bmA$ represents the large-scale part of $\bmA$ 
	by rewriting it in Fourier space. This gives
	\beq
	\bar{\utilde\bmA}(\bmk)=\uGl(\bmk){\utilde\bmA}(\bmk)=e^{-k^2/2k_l^2}{\utilde\bmA}(\bmk).
	\eeq
	
	Since the kernel decreases rapidly for large $k$, as long as the spectrum of ${\utilde\bmA}(\bmk)$ does
	not increase exponentially at large $k$, the spectrum of  $\bar{\utilde\bmA}(\bmk)$
	has little power for $k>k_l$.
%EB11 tweaked above
%EB11 the approximation below best represents the mean if we have a trough  in spectrum near k=1ll  
For $k<k_l$ we  can then write
	\beq
	\uGl(\bmk)=1-\frac{k^2}{2k_l^2}+\mathO\left(\frac{k^4}{k_l^4}\right)+\cdots,
	\eeq
	so that in configuration space $\hg=-\nabla^2/2k_l^2$ [recall the `-' sign in the definition of $\utilde\gamma$ from equation (\ref{gamma2.4})].
	%HZ6
	%EB11 nice
	and $\hg'(A,B)=-\del A\cdot \del B/k_l^2$ for any $A$ and $B$.
	Overall, $\hg$ operating on quantity $Q$ 
	gives $\hg Q\sim (l/l_\text{ch})^2 Q$ where $l_\text{ch}$ is the characteristic variation scale of  $Q$.
	%EB12 thinking out loud, so we need to tweak this further to say that the utility of 2.22 depends on further properties of the spectrum				
					
\subsubsection {\it Moving box average}
	%This is the direct realization of the idea 'averaging over small volumes`. 	
	%In configuration space, 
	Here fields at a point $\bmx$  are averaged in a  finite box with sides of length $l$. 
	Expressing the average using a kernel allows the integral bounds to be taken  to infinity, that is
	%HZ30
	\beq
	\bar\bmA(\bmx)=\frac{1}{l^3}\int_{-l/2}^{l/2}dx'\int_{-l/2}^{l/2}dy'\int_{-l/2}^{l/2}dz' \bmA(\bmx-\bmx')
	=\int d^3x'\ \Gl(\bmx')\bmA(\bmx-\bmx'),
	\eeq
	where $\Gl(\bmx)=\theta_l(x)\theta_l(y)\theta_l(z)$ is the product of three
%	\remark{HZ changed below}
	rectangular functions defined by
	\beq
	\theta_l(x)=\left\{
	\begin{aligned}
		&1/l \ \ -l/2 \leq x \leq l/2 \\
		&0\ \ \text{otherwise}
	\end{aligned}\right\}.
	\eeq
%EB should this be product of three 1-D step functions?
%HZ yes i changed the above.
	We call this a `moving'
	 average because it  is not taken on a fixed grid,
	but centered around each point $\bmx$.   		
	Although suitable for numerical simulation analyses  and seemingly benign, this has  limitations for applicability to realistic contexts.
	 The reason is evident from  the Fourier transform of the kernel of a one-dimensional running box, namely
	\beq
	\uGl(k)=\sinc\left(\frac{kl}{2}\right).
	\eeq
%EB do you mean magntiude  of k or if Kernel is product of three Step functions then dont we get one factor for each, k_x, k_y, k_z....
%HZ i used a 1-D example which i thought was enough for making a point. I have replaced \bm k by k in the eqn. above.
	Here $|\uGl|$ is  a non-monotonic function of $k$ with zero points at $kl=n\pi$, where $n\in \mathbb{Z}$. 
	As a result, some modes with large wave numbers may contribute more to the mean  than those with small wavenumbers.
	%EB11 lets discuss this  ...
	 This  contradicts our  basic notion of mean field theory and highlights
	why  monotonicity of the kernel is a requirement for a physically motivated kernel.  
	A secondary pathology is that  modes with wavelengths $2l/n$ are completely absent from the  mean fields  calculated using this kernel, although  this problem is lessened for large scale  separation  $l\ll \llg$, since then only a few modes lie near  $k=n\pi/l$.
	
 %EB if we shift the intergration bounds then dont we just get the 1-D version of equation 2.22, in which case Kernel is same? I.e. is Kernel in 2.2 product of three step functions in the three directtions then the kernel below is just 1-D of the same?
 %HZ yes
 \subsubsection{Moving line segment  average}
 A one-dimensional, or line average over a segment of length $l$, 
   is a  variant of the moving box  average
    but with the averages  taken   along a single direction $\hr_0$:
    %EB20 minor point: the use of $\hr$ for the line of sight should not be confused with the variable $r$ later used for galactic radius
    %HZ20 changed to $\hr_0$
	\beq
	\bar\bmA(\bmx,\hr_0)=\frac{1}{l}\int^l_0 ds\ \bmA(\bmx+s\hr_0)
	=\int ds\ \theta_l\left(s-\frac{l}{2}\right)\bmA(\bmx+s\hr_0).
	\eeq
Note that the argument of  $\theta_l$ is shifted here because the  line segment over which the average is taken starts from $\bmx$, rather than being centered at $\bmx$.
%HZ i added the kernel and a few words below it.
	The Fourier transform of the kernel is obtained by directly calculating the Fourier transform of $\bar\bmA$, which gives
	\beq
	\uGl(\bmk, \hr_0)=\frac{e^{i\bmk\cdot \hr_0 l}-1}{i\bmk\cdot\hr_0 l}.
	\eeq
	When $|\bmk l|\ll1$, the expansion of $\uGl(\bmk,\hr_0)$ gives $\utilde\gamma=-\frac{i}{2}\bmk\cdot\hr_0 l$ and thus $\hg=\frac{1}{2}l\hr_0\cdot\del$ 
	%HZ6
	and $\hg'=0$. 
	The moving  line segment average has the same
	problems of physical applicability as the moving box average for MFE.
	
	%EB11 add more comment here I think...
	
	\subsubsection{Fixed grid averages}
	We may also average fields inside a set of fixed boxes, i.e., fixed-grid average. 
	For this case, the mean of $\bmA(\bmx)$ is given by
	\beq
	\bar\bmA(\bmx)=\sum_{i=1}^{N_\text{box}} \theta_l(\bmx-\bmx_i)\bar\bmA_{\text{m.b.}}(\bmx_i),
	\eeq
	where $\{\bmx_i\}$, $i=1,2,\dotsb,N_\text{box}$ is the set of points of a grid with side length $l$ and the subscript $\text{m.b.}$ denotes that $\bar\bmA_{\text{m.b.}}$ is a moving box average. 
	Note that since in each grid cell $\bar\bmA(\bmx)$ is a constant, $\bar{\bar\bmA}=\bar\bmA$, this fixed-grid average results in a mean field valued discretely in space. 
	To recover a mean field which is smooth in space, one may apply a second average using a proper  (e.g. Gaussian) kernel. 
	Nevertheless, the averaged field still misses those modes which are resonant to the side length of the grid, and thus   again is physically problematic for observational applications of  MFE.
	
	%HZ4
	\subsubsection{Planar average}
	The planar average is widely used in simulations \citep[e.g.][]{Brandenburg2009,HubbardBrandenburg2011,BhatEbrahimiBlackman2016}.
	It is manifestly  anisotropic since it integrates out, say, $x$ and $y$ but leaves the full $z$ dependence. We write
	\beq
	\bar\bmA(z)=\frac{1}{L^2}\int dxdy\ \bmA(\bmx),
	\eeq
	%EB20 in above equation we change left to be only function of z if L is taken over full area of simulation box?
	%HZ20 yes
	if $L$ is the side length of the simulation box.
	The full Reynolds rules, especially the interchangeability of differential and average operations, are respected if boundary conditions are periodic. That is
	\beq
	\bar{\partial_x \bmA(\bmx)}=\frac{1}{L^2}\int dxdy\ \partial_x \bmA(\bmx)
	=\frac{1}{L^2}\int dy\ \bmA(\bmx)|_{x=0}^{x=L}=0=\partial_x \bar\bmA(\bmx).
	\eeq
	The planar average, although fine for simulation boxes,  does not remove large $k$ modes in the $z$ direction from the mean and so does  not fully filter small scale fields
	from  large scale fields for a real system.
	%\remark{EGB3: tweaked above}
	%EB30 tweaked above
	% and is therefore unable to capture all large-scale features of a realistic system.
	
	\subsubsection {Time average}  
	The time average separates fields into mean and fluctuation components according to their characteristic time variation scales. 
	Mathematically, there is little difference between time average and a one dimensional spatial average if we consider fields to evolve on a four-dimensional space-time manifold, $\bmA=\bmA(t,\bmx)$. 
	The mean quantities are defined via the convolution between the actual fields and a one-dimensional kernel in time, $\bar\bmA(t)=\Gl(t)*\bmA(t)$. 
	For example, a Gaussian kernel is $G_{t_0}(t)=e^{-t^2/2t_0^2}/\sqrt{2\pi t_0^2}$ where $t_0$ is the time scale of average
%	\remark{HZ added below}
	\citep[for applications of space-time filtering, see][]{DakhoulBedford1986a, DakhoulBedford1986b}. 
	%\remark{HZ edited below}
	Optimally practical use such as the minimal-$\tau$ approximation closure \citep{BlackmanField2002}
	%\remark{EGB: shortened here}
	still requires a wide spatial scale separation to ensure
	that temporally averaged quantities decouple from  fluctuating ones.
%to correspond with the 
	%like the ensemble average,
	 %the closure problem 
	 
%HZ4
%EB10 nice, minor tweak below
\subsection{On averages in simulations v.s. observations}
Planar or box averages used in simulations  yield reliable results if compared to a theory based on corresponding averages, and interpreted appropriately. However, it  is  a different question as to how well lessons learned form box averages apply to observations, for which a differently defined mean is more appropriate.
%The only criteria we have used for judging the suitability of kernels is whether large-scale features of fields are correctly captured.

%EB20 am thinking we dont need the table anymore
%\begin{table}
%	\begin{center}
%		\def~{\hphantom{0}}
%		\begin{tabular}{|c|c|c|}
%			{\bf Name}  & {\bf Kernel}   & {\bf Suitable for MFE?} \\ \hline
%			Ensemble   & Distribution function & Y\\
%			Gaussian  & $(k_l^2/2\pi)^{3/2}e^{-k_l^2(\bmx-\bmx')^2/2}$ & Y \\
%			1-D Moving box & $-i\sinc{(k l)}$ & N \\
%			Line-of-sight & $(e^{i\bmk\cdot\hr l}-1)/i\bmk\cdot\hr l$	& N \\
%			Fixed-grid   &  & N \\
%			Temporal   & Any suitable 1-D kernel &  Y 
%		\end{tabular}
%		\caption{
%			Different common averages, their kernels, and whether kernel is suitably monotonic in Fourier space. 
%		}
%		\label{tab_average}
%	\end{center}
%\end{table}

\section{MFE dynamo equations with correction terms}
\label{sec_dynamo}
In this section we re-derive mean field dynamo equations using the kernel formalism of  local averaging
introduced in section \ref{sec_average}, keeping track of correction terms that result from  (i) the non-vanishing means of fluctuations and (ii) the non-equality of double and single-averages. 
With a finite scale separation, these correction terms can be expressed in terms of  mean fields and their spatial derivatives. 
Here we   keep only the  lowest order correction terms, but higher order terms can be derived by the   same method.

\subsection{Derivation}

We average the MHD magnetic induction equation  and use equation (\ref{fml2}), which yields
%HZ6
%HZ30
\beq
\partial_t \bar\bmB
=\del\times\bar{\bm U\times\bm B}
+\nu_{\text{m}}\nabla^2\bar\bmB
=\del\times[(1+\mathO(\hg^2))\bar\bmU\times\bar\bmB-\hg'(\bar\bmU,\bar\bmB)+\emf]
+\nu_{\text{m}}\nabla^2\bar\bmB,
\label{mean_induction_eq}
\eeq
where $\nu_{\text{m}}$ is the magnetic diffusivity assumed to be a constant, and $\emf=\bar{\bmu\times\bmb}$ is the turbulent EMF. 
%[though  $\overline{\bm U}$ does contribute in other terms in equation (\ref{mean_induction_eq})].
%\remark{EGB: edited above}
%we set ourselves in a co-moving frame and drop
 %$\bar\bmU$ hereafter; 
%EB20 inertial frame transformaton does not  kill the gradient of Ubar term, diff rotation.. however, i reworded 
The relative magnitude of  the correction terms to the standard terms 
 that we arrive at in this section are unchanged if $\bar\bmU$ is included.
%EB above statement true and ok?
%HZ Additional terms may occur which are dependent on $\bar\bmU$ but not on $\bar\bmB$; see, for example, \cite{RadlerBrandenburg2010}. The correction of these terms can be derived in the same fashion.

To express $\emf$ in terms of large-scale quantities we adopt the minimal-$\tau$ approach \citep[MTA, ][]{BlackmanField2002}. 
In deriving  $\emf$, terms involving $\bar \bmU$  come proportional to scalar or pseudoscalar cross correlations between functions of $\bm u$ and $\bm b$ \citep[e.g][]{YoshizawaYokoi1993, Blackman2000shear} and for present purposes we ignore these,
%\remark{EB: tweaked here moved EMF sentence here}
%\remark{HZ further edited}
i.e., we ignore terms linear in $\bar \bmU$ or $\bar\bmu$ ($\simeq\hg\bar\bmU=0$) in the evolution equations for $\bm u$ and $\bm b$ (but not in that for $\bar\bmB$).
% since we only use these expressions in the EMF.
The incompressible momentum equation for velocity fluctuations 
%(ignoring contribution from $\bmU$ because they do not contribute to the EMF based
%on our assumption)
then reads
\begin{align}
\partial_t u_l=&
\hat P_{ml}\left[\bar B_n \partial_n b_m+b_n \partial_n \bar B_m+b_n \partial_n b_m
-\bar{b_n \partial_n b_m}\right.\notag\\
&\left.-u_n \partial_n u_m+\bar{u_n \partial_n u_m}
+\hg'(\bar B_n,\partial_n \bar B_m)\right]
+\nu \nabla^2 u_l
\end{align}
where
%\remark{HZ edited below}
$\hat P_{ml}=\delta_{ml}-\partial_m \partial_l\nabla^{-2}$ is the projection operator used to eliminate the sum of thermal and magnetic pressures, and $\nu$ is the viscosity.
%\remark{HZ added below}
The units are such that the mass density $\rho_\text{f}=1$ and the magnetic permeability
%\remark{HZ changed below}
$\mu=1$.
The induction equation for $\bmb$ is
\beq
\partial_t \bmb=\del \times (\bmu\times\bar\bmB+\bmu\times\bmb-\bar{\bmu\times\bmb})
+\nu_{\text{m}}\nabla^2\bmb.
\eeq
%HZ6
Using   equation (\ref{fml1}), and carrying through the  algebra keeping only first order  terms in $\hg$ or $\hg'$, 
we have
%EB12 above tweak ok?
%\remark{HZ removed $\bar{\bmu}$ terms because $\bar{\bmU}=0$ in calculating the EMF}
\begin{align}
\left\{\bar{\bmu\times\partial_t\bmb}\right\}_i
=&\epsilon_{ijk}\left[
(1-\hg)\left(\bar{u_j \partial_n u_{k}}\ \bar{B_n}\right)
-(1-\hg)\left(\bar{u_j u_n}\ \partial_n\bar{B_{k}}\right)
\right.\notag\\
&+\bar {B_n} \hg \bar{u_j \partial_n u_{k}}
-\partial_n\bar{ B_{k} }\hg \bar{u_j u_n}
+\bar{u_j b_n\partial_n u_{k}}
-\bar{u_j u_n \partial_n b_{k}}
%&\left.
%-(1-\hg)\left(\bar{u_j}\ \bar{b_n\partial_n u_{k}}\right)
%+(1-\hg)\left(\bar{ u_j}\ \bar {u_n \partial_n b_{k}}\right)
%-\bar{b_n\partial_n u_{k}}\hg\bar{ u_j}
%+\bar{u_n \partial_n b_{k}}\hg\bar{u_j}
%\right]\notag\\
+\nu_{\text{m}} \epsilon_{ijk}\bar{u_j \partial_{nn} b_k},
\end{align}
and
%HZ7 added the \utilde\gamma' term in the last line in the following
%EB12 yes...i should have noticed before...good
%\begin{align}
%\left\{\bar{\partial_t\bmu\times\bmb}\right\}_i
%=&P_{lj}\epsilon_{ijk}\left[
%(1-\hg)\left(\bar{b_k \partial_n b_{l}}\ \bar{B_n}\right)
%+(1-\hg)\left(\bar{b_k b_n}\ \partial_n\bar {B_{l}}\right)
%\right.\notag\\
%&+\bar {B_n} \hg \bar{b_k \partial_n b_{l}}
%+\partial_n\bar{ B_{l}} \hg \bar{b_k b_n}
%+\bar{b_k b_n \partial_n b_{l}}
%-\bar{b_k u_n \partial_n u_{l}}\notag\\
%&-(1-\hg)\left(\bar {b_k}\ \bar{b_n \partial_n b_{l}}\right)
%+(1-\hg)\left(\bar {b_k} \ \bar{u_n \partial_n u_{l}}\right)
%-\bar{b_n \partial_n b_{l}}\hg\bar {b_k}
%+\bar{u_n \partial_n u_{l}}\hg\bar {b_k}\notag\\
%&\left.
%+\bar{b_k\hg'(\bar B_n,\partial_n \bar B_l)}\right]
%+\nu \epsilon_{ijk}\bar{b_k\partial_{nn} u_j}.
%\end{align}
\begin{align}
\left\{\bar{\partial_t\bmu\times\bmb}\right\}_i
=&\epsilon_{ijk}\bar{b_k \hat P_{lj}(\bar B_n\partial_n b_l+b_n\partial_n \bar B_l)}
+\epsilon_{ijk}\bar{b_k \hat P_{lj}(b_n\partial_n b_l-u_n \partial_n u_l)}\notag\\
&+\epsilon_{ijk}[(1-\hg)(\bar b_k \bar C_j)+\bar C_j \hg \bar b_k]
+\epsilon_{ijk}\bar{b_k \hat P_{lj} \hg' (\bar B_n,\partial_n \bar B_l)}
+\nu \epsilon_{ijk} \bar{b_k\partial_{nn} u_j}
\label{eqn_dtub}
\end{align}
where
%\remark{HZ edited below}
$C_j=\hat P_{lj}({u_n \partial_n u_l}-{b_n \partial_n b_l})$.
Assuming all small-scale quantities are
%\remark{HZ edited below}
isotropic and homogeneous below scale $l$ but could vary on large scales ($\sim\llg$), the two previous equations become
%\remark{HZ changed the sign of the last term in 3.6}
%\remark{HZ edited 3.5 and below}
\begin{align}
\bar{\bmu\times\partial_t\bmb}
=&\left(1-\hg\right)
\left(
-\frac{1}{3}\bar{\bmu\cdot\del\times\bmu}\ \bar\bmB
-\frac{1}{3}\bar{u^2}\del\times\bar\bmB
\right)
+\nu_{\text{m}}\bar{\bmu\times\nabla^2\bmb}
+\bm T^\text{M}\notag\\
&-\hg\left(\frac{1}{3}\bar{\bmu\cdot\del\times\bmu}\right)\bar\bmB
+\hg\left(\frac{1}{3}\bar{u^2}\right)\del\times\bar\bmB,
\label{EMF1}
\end{align}
where
%\remark{HZ changed $T^M$}
$\bm T^\text{M}=\bar{\bmu\times\del\times(\bmu\times\bmb-\bar{\bmu\times\bmb})}$, and
\beq
\bar{\partial_t \bmu\times\bmb}
=\left(1-\hg\right)\left(\frac{1}{3}\bar{\bmb\cdot\del\times\bmb}\ \bar\bmB\right)
+\hg\left(\frac{1}{3}\bar{\bmb\cdot\del\times\bmb}\right)\bar\bmB
+\nu\bar{\nabla^2\bmu\times\bmb}
+\bm T^U,
\label{EMF2}
\eeq
where $T^U_i=\epsilon_{ijk}\bar{b_k \hat P_{lj}(b_n\partial_n b_l-u_n \partial_n u_l)}$.
%\remark{HZ added below}
The derivation of the first and second terms in equation (\ref{EMF2}) is given in appendix \ref{appx2}.
Also note that the small-scale part in the $\hg'(\bar B_n,\partial_n\bar B_l)$ term will have its maximum wavenumber at $\sim 2\klg$.
Therefore if the scale separation is large enough such that $2\klg\ll\ksm$, the $\hg'(\bar B_n,\partial_n\bar B_l)$ term can be roughly treated as a large-scale quantity in equation (\ref{eqn_dtub}).

Adding equations (\ref{EMF1}) and (\ref{EMF2})  gives
%\remark{HZ edited below to include the last two terms}
\begin{align}
\partial_t \emf
=&(1-\hg)\left(\tilde\alpha\bar\bmB-\tilde \beta\del\times\bar\bmB\right)
+(\hg\tilde\alpha)\bar\bmB-(\hg\tilde\beta)\del\times\bar\bmB\notag\\
&+\nu_{\text{m}}\bar{\bmu\times\nabla^2\bmb}+\nu\bar{\nabla^2 \bmu\times\bmb}
+\bm T^M+\bm T^U
\label{eqn_evol_emf}
\end{align}
where 
$\tilde \alpha=(\bar{\bmb\cdot\del\times\bmb}-\bar{\bmu\cdot\del\times\bmu})/3$ and $\tilde \beta=\bar{u^2}/3$.
In the spirit of the MTA, 
the sum of the triple correlation terms in equation (\ref{eqn_evol_emf}) is equated to a damping term $-\emf/\tau$. 
%(We note that the correction in \cite{Rheinhardt2012} amounts to adding a spatial damping term to the  temporal damping term)
For  $|\tau \partial_t \emf|\ll|\emf|$,
%the turbulent EMF generally saturates much faster than $\bar\bmB$, quasi-stationary condition
equation (\ref{eqn_evol_emf}) then gives
\beq
\emf=(1-\hg)\left(\alpha \bar\bmB-\beta\del\times\bar\bmB\right)
+(\hg\alpha)\bar\bmB-(\hg\beta)\del\times\bar\bmB
\label{emf}
\eeq
in the ideal MHD limit $\nu,\nu_\text{m}\to0$, 
where $\alpha =\tau \tilde \alpha$ and $\beta =\tau\tilde \beta$  are the  
helical and diffusion dynamo coefficients  
and
%\remark{HZ edited below}
$\tau$ is %best interpreted as 
the damping time for the EMF when mean 
fields are removed.  Empirically, this is  approximately equal to the turnover time at the turbulent
driving scales in forced isotropic simulations \citep{BrandenburgSubramanian2005MTA}.
We also define $\alphak=-\tau\bar{\bmu\cdot\del\times\bmu}/3$ and $\alpham=\tau\bar{\bmb\cdot\del\times\bmb}/3$ being the kinetic and magnetic contributions to the $\alpha$-effect, respectively.

When there is large scale separation 
 $\hg,\hg'\to0$, equations (\ref{mean_induction_eq}) and (\ref{emf}) reduce
exactly to the standard dynamo equations derived with ensemble average. 
This important feature indicates that different kinds of suitable averaging - like local Gaussian average or ensemble average - converge to the same set of equations when scale separation is large.
%EB in the previous section, we said that moving box, line of sight are not suitable so this confusing perhaps to mention here without further discussion.
%HZ deleted those two averages. 

%\remark{HZ added below, EB tweaked}
The turbulent EMF now has  routes of expansion:  (i)  higher gradients of $\bar{\bmB}$; (ii)  $\hg$ due to the violation of Reynolds rules.
Expanding to every higher order results in (using order-of-magnitude estimates) an extra factor of $\lsm/\llg$ for the former, and $(l/\llg)^c$ for the later.
Interestingly, both of these two ratios  are related to the scale separation, and the question of which dominates higher order terms in $\emf$ varies for different models.
In this work we assume the $\hg$ corrections 
%always 
dominate.

\subsection{Comparison to previous work on non-local EMF kernels}
%EB4 edited section a bit
We see from equation (\ref{emf}) that the violation of the Reynolds rules from mesoscale fluctuations is a direct source of  contributions to the EMF from   terms with higher than linear order in derivatives of $\bar\bmB$.
%EB4 thinking out loud: in older work like moffatts book,  there is an anzatz expansion of Bbar in terms of all higher derivatives. But when reynolds rules hold,  these terms must come from the closure because if one ignores the triple correlations, there are no such terms/
In Fourier space these terms imply that  $\utilde{\mathcal{E}}_i(\bmk)=\utilde K_{ij}(\bmk)\utilde{\bar B}_j(\bmk)$ where $\utilde K_{ij}$ could contain terms of order higher than linear in 
%HZ30
%$|\bmk|$
$\bmk$, in contrast  to the conventional  mean field dynamo theory where $\utilde K_{ij}=\alpha\delta_{ij}-i\ep_{imj}\beta k_m$.
%\remark{EGB: deleted rest of sentence.}
%and higher order terms that might otherwise arise separately from  the closure are assumed to be small \citep{Moffatt1978}.  

%EB20 people have expanded the EMF in terms of higher order derivatives of Bbar in the past but i think the extra terms coming from the clousure. Also Brandnebrug and Sarson 2002
Consequently, in configuration space we have $\emf(\bmx)=\bm K*\bar\bmB$, and the turbulent EMF depends on $\bar\bmB$ through its weighted average in the vicinity of $\bmx$, i.e., non-locally.
More generally, if we have used a time average in equation (\ref{emf}), $\bmK$ could also be time-dependent, and correspondingly $\emf$ becomes non-local in both space and time.
 
 The EMF kernel  $\bmK$ that we derive
% \remark{HZ edited below}
 includes terms caused by 
% is directly determined  by 
violation of Reynolds rules, 
 and  varies depending on the choice of our (potentially anisotropic) averaging kernel $\bm G$. 
Although previous work  has  identified  the need for an EMF kernel to capture non-locality \citep{KrauseRadler1980, Radler2000, RadlerRheinhardt2007, Brandenburg2008, HubbardBrandenburg2009, Rheinhardt2012}, this previous work did not
address the  contribution to this kernel from  the violation of the Reynolds rules.
%\remark{EGB: deleted below} 
%but instead is implicitly addressing the closure.
For example,   \cite{Rheinhardt2012} used numerical simulations (DNS) to test an anzatz
for the EMF kernel in the case of 
homogeneous isotropic turbulence with mean fields defined by an average over the $x-y$ plane.
They assessed whether the EMF kernel takes the form
\beq
\utilde K_{ij}(\omega,\bmk)=\frac{\alpha\delta_{ij}-i\ep_{imj}\beta k_m}{1-i\omega\tau_{\text{RB}}+l_{\text{RB}}^2 k^2}
\label{eqn_RB_kernel}
\eeq
at low wavenumbers, where $\tau_{\text{RB}}$ is approximately equal to the eddy turnover time $\tau$, and $l_{\text{RB}}$ is a parameter whose value is to be extracted  from  fits to simulation data.
Their resulting evolution equation for the EMF reads
\beq
\left(1+\tau_{\text{RB}}\partial_t-l_{\text{RB}}^2\partial_z^2\right)\emf
=\alpha\bar\bmB^{(\text p)}-\beta\del\times\bar\bmB^{(\text p)},
\label{eqn_RB_evol_emf}
\eeq
where the superscript `$(\text p)$' distinguishes their planar average from our kernel averages.
%EB4 thinkng out loud:  there is a kernel for planar average
\cite{Rheinhardt2012} found that equation (\ref{eqn_RB_kernel}) was  at least  consistent
with simulation data  up to $k/k_1\approx3$ where $k_1$ is the simulation box wavenumber,
for $\tau_{\text{RB}}\sim \tau $ and $l_{\text{RB}}\sim \lsm$, the  energy-dominating  eddy scale. 
%EB20 tweaked above

To compare  equation (\ref{eqn_RB_evol_emf}) with our result equation (\ref{eqn_evol_emf}), we ignore the second to fifth terms on the RHS of the latter  (i.e., assuming $\alpha$ and $\beta$ are constants and taking the ideal MHD limit), and  identify the sixth and seventh terms (triple correlations) with $\hat T \emf$ where $\hat T$ is an operator.
This gives
\beq
(-\tau\hat T+\tau \partial_t)\emf=(1-\hg)\left(\alpha\bar\bmB-\beta\del\times\bar\bmB\right).
\label{eqn_evol_emf_2}
\eeq
The identification of the triple correlations with a damping term, $\hat T=-1/\tau$, serves as the closure in MTA.
%Motivated by \cite{Rheinhardt2012}, our new understanding of
 Comparison to equation (\ref{eqn_RB_evol_emf}) shows that 
  %closure to include the third term on the LHS of equation (\ref{eqn_RB_evol_emf}), namely, in the evolution equations of $\emf$, 
  the left sides of the two equations can be made to mutually correspond if
  we    replace the triple correlations by the sum of a damping term and a diffusion term, that is 
  $\hat T \emf=-\emf/\tau+\eta_\text{t.c.}\nabla^2\emf$, where $\eta_\text{t.c.}=l_{\text{RB}}^2/\tau$ is a diffusion coefficient determined by statistical properties of turbulent fields.  
The spatially non-local term in equation (\ref{eqn_RB_evol_emf}), $-l_{\text{RB}}^2\partial_z^2\emf$, can thus be  understood as textured specification
of  the form of  terms for which the crude MTA approximates. This additional term 
 plays a similar  role to that of the standard MTA term, namely that it  depletes 
 the turbulent EMF in the absence of any other mean fields.  
% While the long term  goal of identifying separate fit parameters 
 %$l_{\text{RB}}$ and $\tau_{\text{RB}}$ from simulations is 
%It is  not yet clear that the use of  two fit parameters $l_{\text{RB}}$ and $\tau_{\text{RB}}$  determined empirically from simulations is more theoretically robust than  MTA with a single parameter.
 %The diffusion term emerges from closure and therefore the potential difficulty of its application lies in the fact that fitting parameters like $l_{RB}$ still lack theoretical predictions and can only be determined via simulations.

 We emphasize that the derivation of
 equation (\ref{eqn_evol_emf_2})  differs from that of  equation (\ref{eqn_RB_evol_emf}) in  that  the correction terms  appearing on the right  of equation (\ref{eqn_evol_emf_2}) 
 are derived from  the averaging procedure itself, and represent the lowest order corrections when Reynolds rules are violated. Higher order terms can also be derived. 
The form of $\hg$ is determined by the scale $l$ and the kernel of average.
 These terms are not included in the semi-empirical approach of \cite{Rheinhardt2012} that produced Eq. (\ref{eqn_RB_evol_emf}),
% \remark{HZ edited below}
 because they vanish identically due to the planar average. 
 %equation (\ref{eqn_RB_evol_emf}) suggests an additional correction of traditional mean field dynamo theory, independent of those rise from average procedure as in equation (\ref{eqn_evol_emf_2}). 
%Besides the physical meanings of their correction terms, 
%equation (\ref{eqn_evol_emf_2}) also differs from equation (\ref{eqn_RB_evol_emf}) in two more aspects.
 %Moreover  equation (\ref{eqn_RB_evol_emf}) is 

Finally, we note that in deriving equation (\ref{eqn_evol_emf_2}), we  averaged the MHD equations using a kernel that retains a spatial dependence, 
%configuration and Fourier space,
 so that  $\bar\bmB$ can depict large-scale magnetic fields and retain large scale gradients in all directions.
%EB4 
In contrast, 
  the $x-y$ planar average used in \cite{Rheinhardt2012} does not retain large scale field gradients in $x$ and $y$ directions, which is self-consistent for the simulation boxes but not sufficiently general for investigating mean fields. 
In addition  planar averages do not remove  large $k_z$ modes from
$\bar\bmB^{(\text p)}$, and hence equation (\ref{eqn_RB_evol_emf}) might not be  complete even for the simulations in the absence of  including higher order terms since  the EMF kernel (\ref{eqn_RB_kernel}) is valid only for small $|\bmk|$. 
%while $\bar\bmB^{(p)}$ contains large wavenumber modes.
%This could be overcome by choosing a  different  averaging kernel that retains this spatial dependence. 

\section{Precision of mean field theories}
\label{sec_twoerrors}
The precision error of a mean field theory (MFT) can be classified into two types:
(i)  intrinsic error (IE) $\errorint$ from the theory itself and 
(ii)  filtering error (FE) $\errordf$ associated with comparing the mean field theory values filtered through a measuring kernel (thus double-filtered) 
  with the total field filtered through the measuring kernel.   Both of these  depend on the scale of average $l$. We now derive these in full.

\subsection{Intrinsic error from  statistical fluctuations in inputs to mean field equations}
%\remark{HZ edited below}
%\remark{HZ edited Sec. 4.1; EGB: edited further}
The dynamo  input parameters in the mean field equations
[e.g., $\alpha$ and $\beta$ in equation (\ref{emf}) along with boundary and initial conditions]
are themselves random variables (in an ensemble) and so is $\bar\bmB=\bar\bmB(\bmx;\alpha,\beta,\cdots)$,
%Cells of scale $\lsm$ 
%contributing to the mean (of scale $l$)  can incur statistical fluctuations in the turbulent quantities about the means  that define $\alphak$ and $\beta$.
because the small-scale fields $\bmu$ and $\bmb$ are statistically fluctuating.
The intrinsic error is thus defined as the variation of statistical fluctuations of $\bar{\bmB}$ (about its ensemble mean) due to these small-scale fluctuations, which we denote by $\errorint$:
\beq
\sigma^2_{\text{IE},B_i}=\langle{(\bar{B}_i-\langle{\bar{B}_i}\rangle)^2}\rangle,\ \text{for }i=1,2,3.
\eeq
With this definition the IE vanishes if the mean field theory is defined using ensemble average, i.e., $\sigma^2_{\text{IE},B_i}=\lra{(\lra{B_i}-\lra{\lra{B_i}})^2}=0$.

The IE can be calculated by propagating the statistical variations of input parameters to the solutions of mean field equations.
%The  specific forms varying for different models. 
%EB20 tweaked above
We consider the IE of  the steady-state solutions of MFE dynamo equations for a minimalist model where  $\alphak$ and $\beta$ are the only input parameters: $\bar\bmB=\bar\bmB(\bmx; \alphak,\beta)$.
The magnetic $\alpha$-effect, $\alpham$, is  dynamical in our model, and not an input parameter,  since it is governed by the transport equation of the helicity density, equation (\ref{eqn_helicity}).

Let us consider a minimalist model where all 
%un-averaged 
%\remark{EGB: later we want $\alphak$ to be "unaveraged"--
%so I put a "bar" over $\alphak$ in its use below--correct?}
%\remark{HZ: removed bar there}
turbulent transport coefficients 
are statistically homogeneous over the whole space.
%\remark{HZ removed the sentence below}
%for example, $\alphak=\bar{\tau\bmu\cdot\del\times\bmu/3}$ is independent of $\bmx$ for sufficiently large $l$, and equal to its ensemble averaged value $\alpha_{\text{k}0}=\langle\tau\bmu\cdot\del\times\bmu/3\rangle$.
%For small $l$ (but still $l>\lsm$), $\alphak$ is spatial dependent.
In one such dynamo model that we discuss later, turbulent transport coefficients depend on the radial coordinate, but since its variation scale is greater than $l$,  they remain locally approximately homogeneous.
%In fact l should never go beyond the "large scales" of any quantities.

%If ensemble averaging is replaced by filtering,
The deviation of kernel-filtered values from the ensemble averages of turbulent coefficients contributes to the IE of the mean fields.
The resulting average (in the sense of an ensemble average) imprecision in $\bar{\bm B}$ can be calculated by propagating the imprecision to turbulent coefficients.
For $\alphak$, this is
%\remark{HZ edited below}
\beq
\sgma^2=\langle(\alphak-\lra{\alphak})^2\rangle
%=\langle\alphak^2\rangle-\alpha_{\text{k}0}^2.
\label{eqn_4.1a}
\eeq
Similarly, we have
\beq
\sgmb^2=\lra{(\beta-\lra{\beta})^2},
\label{eqn_4.1b}
\eeq
and
\beq
\sgmab^2=\langle(\alphak-\lra{\alphak})(\beta-\lra{\beta})\rangle.
\label{eqn_4.1c}
\eeq
The  uncertainty in $\bar\bmB$  derives from  the  uncertainties 
from  $\alphak$ and $\beta$  as follows:
\beq
\errorintb
=(\partial_{\alphak}\bar B_i)^2 \sgma^2
+(\partial_\beta \bar B_i)^2 \sgmb^2
+2(\partial_{\alphak}\bar B_i)(\partial_\beta \bar B_i) \sgmab.
\label{errorint_of_Bbar}
\eeq
%where $\sgma$, $\sgmb$, and $\sgmab$ are given by equations (\ref{eqn_4.1a}) to (\ref{eqn_4.1c}).
%which could be reduced be magnetic helicity conservation
%EB12 not sure if it could be elimiated by helicity conservation as we have a helicity flux..

%$\alpha_m$ 
%could be eliminated via helicity conservation.
%\beq
%\sgma^2=\frac{1}{9}\bar{(\tau\bmu\cdot\del\times\bmu)^2}-\alphak^2,
%\ \ 
%\sgmb^2=\frac{1}{9}\bar{\tau^2 u^4}-\beta^2,
%\eeq
%and 
%\beq
%\sgmab=\bar{
%	\left(-\frac{1}{3}\tau\bmu\cdot\del\times\bmu-\alphak\right)
%	\left(\frac{1}{3}\tau u^2-\beta\right)
%}.
%\eeq
%Note that $\sgma^2$ and $\sgmb^2$ represent variances of the  turbulent products
% $-\tau \bmu\cdot\del\times\bmu/3$ and $\tau u^2/3$ respectively,
%and lead to local variations about the means, not variation of the means  $\alphak$ and $\beta$ themselves.
%EB20 above tweak ok?
%HZ20 yes thanks
%which is the mean of $-\tau \bmu\cdot\del\times\bmu/3$ ($\tau u^2/3$).

%For present purposes, we restrict attention  to $\sigma_{\alpha_k}$ and
% $\sigma_{\beta}$  which we will combine in quadrature along with equation (\ref{var_result})
%to get a lower limit on the imprecision of $\bar\bmB$.

To estimate magnitudes of equations (\ref{eqn_4.1a}) to (\ref{eqn_4.1c}),
%\remark{HZ edited below}
we decompose filtered quantities into (ensemble averaged) means and random parts.
%\remark{EGB: There is some ambiguity /contradiction with previous page as to whether $\alpha_k$ should be averaged or unaveraged}
Consequently we have
%\remark{EB lets chat on eqn below}
\beq
\alpha_\text{k,r}=\alphak-\lra{\alphak}=\frac{\tau}{3}\bar{\bmu\cdot\del\times\bmu-\lra{\bmu\cdot\del\times\bmu}},
\label{eqn_4.1decomp}
\eeq
where $\alpha_\text{k,r}$ is the random part.
%and coherent in each cell of size $\lsm/2$ with 2 accounting for it being quadratic in $u$ whose coherent scale is $\lsm$.
Combining equations (\ref{eqn_4.1a}) and (\ref{eqn_4.1decomp}) we have
\beq
\sgma^2=\langle\alpha_\text{k,r}^2\rangle.
\label{eqn_4.1ar}
\eeq
Similarly,
\beq
\sgmb^2=\langle\beta_\text{r}^2\rangle
\label{eqn_4.1br}
\eeq
where 
\beq
\beta_\text{r}=\beta-\lra{\beta}=\frac{\tau}{3}\bar{u^2-\lra{u^2}}
\label{eqn_4.1decompbeta}
\eeq
is the random part of $\beta$.

To estimate the quantities in
 %$\bar \alpha_{\text{r}}$ and $\bar \beta_{\text{r}}$  in 
 equations
  (\ref{eqn_4.1ar}) and (\ref{eqn_4.1br}), 
%using mean quantities, 
%the contributions to the IE
we consider  the system  of study to be divided
into cells of typical length $\lsm$ 
and crudely assume that  in each cell, $\bmu$ is nearly uniform with components  drawn from independent Gaussian distributions,
\beq
f(u_i)=\frac{1}{\sqrt{2\pi}u_0}e^{-u_i^2/2u_0^2},\ i=1,2,3.
\eeq
Then for each cell,
\beq
\lra{u_i}=0,\ 
\lra{u_i^2}=u_0^2,\ 
\lra{u_i^4}=3u_0^4,
\eeq
and
\beq
\lra{u^2}=\sum_{i=1}^3 \lra{u_i^2}=3u_0^2,\ 
\lra{u^4}
=\lra{\left(\sum_{i=1}^3 u_i^2\right)^2}
=15u_0^4
=\frac{5}{3}\lra{u^2}^2,
\eeq
so that
\beq
\sigma_{u^2}^2=\lra{u^4}-\lra{u^2}^2=\frac{2}{3}\lra{u^2}^2
\label{eqn_4.1var}
\eeq
which links fluctuations to mean quantities.

The filtering $\bar{(\cdot)}$ in equations (\ref{eqn_4.1ar}) and (\ref{eqn_4.1br}) can be roughly seen as the algebraic average of the quantity $(\cdot)$ of $N=(2l/\lsm)^3$ cells, with the factor of two accounting for the fact that the variation scale of $u^2$ will be $\lsm/2$ if that of $\bm u$ is $\lsm$.
The central limit theorem (CLT) then yields
\beq
\sgma^2\simeq\frac{\langle\alpha_{\text{k,r}}^2\rangle}{N}
=\frac{\sigma^2_{\alpha_\text{k,r}}}{N},\ 
\sgmb^2\simeq\frac{\langle\beta_{\text{r}}^2\rangle}{N}
=\frac{\sigma^2_{\beta_\text{r}}}{N},
\label{eqn_4.1clt}
\eeq
where $\sigma^2_{\alpha_\text{k,r}}$ and $\sigma^2_{\beta_\text{r}}$ are the variances of the random parts in each cell.

Since both $\alphak$ and $\beta$ are quadratic in $\bmu$, equations (\ref{eqn_4.1var}) and (\ref{eqn_4.1clt}) then yield
%allows us to compute  fluctuations about those means  using the central limit theorem (CLT), by dividing by $N$,
%where 
%$N= (2l/\lsm)^3$ is the number of cells in the region of average $l^3$.
%The factor of two is due to the fact that the variation scale of $u^2$ will be $\lsm/2$ if that of $\bmu$ is $\lsm$.
%The result is then 
\beq
\sgma^2\simeq\frac{2\alphak^2/3}{(2l/\lsm)^3}, \ \ \ 
\sgmb^2\simeq\frac{2\beta^2/3}{(2l/\lsm)^3}.
\label{sigma_a_and_b}
\eeq
It then follows from equation (\ref{errorint_of_Bbar}) that
\beq
\errorintb
\simeq\frac{1}{12(l/\lsm)^3}\left[
(\partial_{\alphak}\bar B_i)^2 \alphak^2
+(\partial_\beta \bar B_i)^2 \beta^2
\right]
+2(\partial_{\alphak}\bar B_i)(\partial_\beta \bar B_i) \sgmab.
\label{errorintbi}
\eeq
Note that $\errorintb$ depends on spatial coordinates $\bmx$, just as $\bar\bmB$ does.
In galaxies, a typical value of the variation scale of turbulent fields satisfies $\lsm\lesssim0.1\ \text{kpc}$.
Hence  $(l/\lsm)^3\gtrsim8$ for $l=0.2\ \text{kpc}$ and $\gtrsim64$ for $l=0.4\ \text{kpc}$.

From the CLT, this IE decreases with increasing $l$   because  the average variations from turbulence  are inversely proportional to the number of eddy cells in the region being averaged, $(l/\lsm)^3$, provided that $\lsm$ is rather insensitive to the choice of $l$.

\subsection{Filtering error  from  mismatch between measurement and theoretical kernels}
Measuring physical quantities always results in measuring mean quantities to a certain extent. Detectors    have limited sensitivity  so measurements represent a convolution between true physical quantities and an instrument kernel. Furthermore, the physical quantity being measured
typically involves a superposition of microphysical contributions and an  average over many  local  macroscopic contributions.
In particular, the predicted values of observed FR and synchrotron polarization are limited in precision when these predictions are made using MFE.
% a physical mean, such as a line of sight average.
%Fnite resolutions of apparatus unavoidably lead to the incapability of measuring physical quantities at a single spatial point and instant time; 
%instead, objects are blurred spatially and temporally, and measurements give the in accordance to observational instruments being used. 
%Secondly, sometimes observational techniques themselves intrinsically carry averages. 
%For example, two important phenomenon for measuring astrophysical magnetic fields, 
%averages along the line of sight 
%In short,  MFE  allows us to  makes predictions for mean quantities
% with a physical averaging built into the theory.  

%\remark{HZ edited below}
%\remark{EB   edited paragraph}
For a  given physical quantity of a real system $Q^\text{A}$  (e.g., the actual magnetic field of a galaxy) 
we define the measured value as  $(Q^\text{A})_\mathcal{M}$,  where the subscript $\mathcal{M}$ indicates
that quantity subjected to a measuring  kernel that the instrument
uses to project out the actual measured value.
Complementarily, we write $\overline{Q^\text{A}}$ to  indicate 
the value of $Q^\text{A}$ subjected to a theoretically chosen mean field theory filter.
We will assume these two filters commute, i.e., $\overline{(Q^\text{A})_\mathcal{M}}=(\overline{ {Q^\text{A}}})_\mathcal{M}$.
%\remark{HZ: regarding the accuracy issue:}
We use $Q$ to indicate  a theoretically predicted value of $Q^\text{A}$.
Like $Q^\text{A}$ we can subject $Q$ mathematically to a theoretical mean field filtering and obtain $\bar Q$ or to measurement filtering to obtain $(Q)_\mathcal{M}$, or both $({\overline Q})_\mathcal{M}$ ($= \overline{(Q)_\mathcal{M}}$ by assumption).
%The observed  measurement  $(Q^\text{A})_\mathcal{M}$, and  the theoretical prediction as 
%$(\bar{Q})_\mathcal{M}$.
%The latter quantity is subjected to $\mathcal{T}$-filtering to indicate predicted value from mean field theory.
For the common practice in which observations are not subjected to the theoretical mean filtering but the theory is subjected to the instrument filtering,  the difference  the measured value and the theoretically predicted mean can be written
\begin{align}
(Q^\text{A})_\mathcal{M}-(\bar{Q})_\mathcal{M}
&=[{\bar{(Q^\text{A})}_\mathcal{M}}-(\bar{Q})_\mathcal{M}
+(q^\text{A})_\mathcal{M}-(q)_\mathcal{M}]
+(q)_\mathcal{M}\notag\\
&=[(\bar{Q^\text{A}}-\bar{Q})_\mathcal{M}
+(q^\text{A}-q)_\mathcal{M}]
+(q)_\mathcal{M}
\label{eqn_QOQTH}
\end{align}
where
%\remark{EB tweaked what below to make more explicit for posterity}
 $(q^\text{A})_\mathcal{M}=(Q^\text{A})_\mathcal{M}-{\bar{(Q^\text{A})}_\mathcal{M}}$ is the difference between the actual quantity and its value using the  theoretical  mean filter, then both
filtered through the measuring kernel. Analogously,  $(q)_\mathcal{M}= (Q)_\mathcal{M}-(\bar{Q})_\mathcal{M}$
is  the difference between the theoretical quantity and its theoretical mean filter 
then both filtered through the  measuring kernel.
The terms in the square brackets on the right of  equation (\ref{eqn_QOQTH}), 
 measure {\it accuracy} of the theoretical model and these terms will be  small  
%(i.e., how well the theory fits the actual system)
 if the theory provides a good match to the real system.
We focus on  $(q)_\mathcal{M}$,  the last term in equation (\ref{eqn_QOQTH}),  which is a precision error of the theory and the  FE that we will quantify. The smaller its magnitude, the more precise the theory.
%In what follows, we focus on quantifying and minimizing $(q^{\text{TH}})_\mathcal{M}$.

In principle, 
  one would like to subject  $(Q^\text{A})_\mathcal{M}$  to the same filtering which corresponds to that of the mean field model, that is,  compute $\bar{(Q^\text{A})_\mathcal{M}}$,
and compare it to $(\bar{Q})_\mathcal{M}$. 
This would obviate computation of the FE.
%only then compare parison between theories and observations
 For simulations this may be possible, but 
for observations  one cannot always compute $\bar{(Q^\text{A})_\mathcal{M}}$, due to limited resolution. Moreover, it is typically not  done in practice, and cannot be done if $\bar{(\cdot)}$ represents the ensemble average and the system
 has finite scale separation. 
 If both $\bar{(\cdot)}$ and $(\cdot)_\mathcal{M}$ averages were  equivalent to ensemble averages
due to infinite scale separation, then  $(q)_\mathcal{M}=\lra{Q-\lra{Q}}\rightarrow 0$; 
%.e.
% no deviation between theoretical prediction and measurement  would be expected.  
 but this is not the case with finite scale separation and local spatial averages.
 %, where fluctuations have non-vanishing mean values regardless of whether $\mathcal{T}$ and $\mathcal{M}$ are the same or not.

Unlike the IE of the previous subsection, the FE {\it increases} with increasing $l$, since smaller $l$ means including a greater fraction of modes into what comprises the mean field, and so the  theoretical predictions from mean field theory would be less coarse-grained and  thus more  capable of characterizing the actual field.
%\remark{EGB: commented out a few sentences that seem to be unnecessary here}
%In addition, the IE of $(Q_\mathcal{T})_\mathcal{M}$ is given by propagating the IE of $Q_\mathcal{T}$, $\sigma^2_{\text{IE},Q_\mathcal{T}}$.
%If the correlation lengths of fluctuation fields are small compared to the scale of average of $\mathcal{M}$, then this IE of $(Q_\mathcal{T})_\mathcal{M}$ is roughly equal to 
%$(\sigma^2_{\text{IE},Q_\mathcal{T}})_\mathcal{M}$.
If the presumption is made that IE and FE are statistically independent and uncorrelated for a given $l$, then the total uncertainty of the mean field theory is given by $\sigma^2=\errorint+\errordf$.
Due to their competitive behaviors when changing $l$, an optimal scale of average $\lopt$ which minimizes either $\sigma^2$ or the relative uncertainty $\sigma^2/\bar B^2$ can arise, satisfying $\lsm<\lopt<\llg$.

In the next section we combine all of the formalism of this  section into a specific example.
We  discuss implications of a finite precision when comparing
observations and MFE theory for measuring galactic fields by FR  from extragalactic sources. Our formalism is not restricted to that particular example
and the precision of theoretical predictions for other kinds of observations, such as pulsar FRs, or polarized synchrotron emission, can also similarly be worked out.

\section{MFE precision error in the context of  FR}
\label{sec_precision}
FR is  commonly used to measure   strengths and directions of magnetic fields in galaxies. 
The rotation measure (RM), i.e., the rotation of the polarization plane of light from a distant pulsar or extragalactic radio source is given by \cite{RSS1988} to be
%\remark{HZ edited below}
\beq
RM=0.81\int d\bm s \cdot \bmB n_e\ (\text{rad\ m}^{-2})\ 
\propto \int d\bm s \cdot \bmB,
\eeq        
where the integrals are along the line of sight, and the  proportionality is
valid when the thermal electron density $n_e$ varies on scales larger than  those of $\bmB$.
% in  the integration region.
%HZ add n_e.. ref: shukurov review

Here we focus on the RMs  through a  galaxy other than the Milky Way 
from extragalactic sources, and leave the discussion of pulsar RMs in the Milky way for section \ref{sec_measure_in_galaxy}.  We  also omit any  influence of weak intergalactic magnetic fields. 
%Due to the weakness of of magnetic fields outside the galaxy,
 The relevant segment of integration is then   the segment of each line of  sight $L(R,r,h)$  inside the  galaxy  [see  figure \ref{schematic_plots}  with an (a) edge-on view, (b) face-on view, and (c) inclined view], where
$L$ is a function of the galactic radius $R$, the distance $r$ from the line of sight to the galactic center, and the semi-thickness of the galactic disk $h$.

In what follows, we use 
%\remark{EGB: we somewhat change the meaning of subscript $L$ from previous sections}
%We introduce 
the subscript $L$\footnote{
This shall not be confused with, say, the characteristic length of large-scale quantities $\llg$, whose subscript is in roman type.	
}
for a constant thermal electron density 
FR-like average 
 along path $L$. For a given vector field $\bm Q$ and scalar field $f$
this line of sight average gives
%\remark{HZ edited below:EGB: tweaked paragraph below further}
\beq
(Q)_L=\frac{1}{L}\int d\bm s \cdot \bm Q  \ {\rm and}\  (f)_L=\frac{1}{L}\int ds\ f.
\label{eqn_los_average}
\eeq
For FR measurements the line-of-sight average $(\cdot)_L$ will thus correspond to $(\cdot)_\mathcal{M}$ mentioned above.
We denote the theoretical prediction of the line-of-sight mean field from MFE  as $(\bar B)_L$.
While $\errorint$ of $(\bar B)_L$ can be computed by propagating the IE  of $\bar\bmB$,
the FE $\errordf$ arises from calculating the difference $(b)_L$, between $(\bar B)_L$ and $(B)_L$, the latter determined by how the RMs are measured. We have
\beq
(b)_L
\equiv (B)_L- ({\bar B})_L
=\frac{1}{L}\int d\bm s\cdot \bmB-\frac{1}{L}\int d\bm s\cdot\bar\bmB
=\frac{1}{L}\int d\bm s\cdot \bmb.
\label{b_L}
\eeq
This represents the line-of-sight mean of a fluctuation 
and is the deviation that results from comparing single average to a mixed 
 double average $(q)_\mathcal{M}$, discussed in section \ref{sec_twoerrors}. 
% It is the FR-like average of the fluctuation of $\bmB$ associated with whatever kernel  is used to define the mean field theory.
Here $(\bar{Q})_\mathcal{M}=({\bar B})_L=\frac{1}{L}\int d\bm s\cdot\bar\bmB$.

Our mission is to express $\errorint$ and $\errordf=\lra{(b)_L^2}$ in terms of known or derivable quantities for a MFT.
Equation (\ref{errorintbi}) gives the general form of $\errorint$ for  $\bar\bmB$.
If fluctuations in different directions are uncorrelated, the intrinsic error of $(\bar B)_L$ can  be approximated by
\beq
\errorint
\simeq
\frac{1}{L}\int ds\ 
\left[
(\sigma_{\bar B_x}\hat{\bm x}\cdot \hat{\bm s})^2
+(\sigma_{\bar B_y}\hat{\bm y}\cdot \hat{\bm s})^2
+(\sigma_{\bar B_z}\hat{\bm z}\cdot \hat{\bm s})^2
\right],
\label{errorint}
\eeq
 where fluctuation scales are less than $L$, which is true away from the galactic edge.
%EB20 tweaked above--toward the edge i guess where L~ l it breaks down a alittle.
%HZ20 yes

To  compute $\lra{(b)_L^2}$ we assume a statistically isotropic turbulent field $\bmb(\bmx)$, and therefore the integrand on the RHS in equation (\ref{b_L}) is insensitive to the direction of the line of sight. 
%\remark{HZ edited below}
We use a scalar $b_\text{s}(x)$ to represent the component of $\bmb(\bmx)$ along the line of sight. 
Next, we assume $b_\text{s}(x)$ can be decomposed into different modes with specific  wavelengths indicated by a superscript $(m)$,  namely
%EB2 do we want to keep vector in argument of b rather than scalar?
%HZ2 I think we don't need to because of istropy.
\beq
b_\text{s}(x)=\sum_m b^{(m)}(x),
\eeq
with $k_m$ being the characteristic wavenumber of each mode and satisfying
\beq
\frac{2\pi}{k_m}\leq l, 
\eeq
since the turbulent scale is smaller than the averaging scale.
%\remark{HZ edited for the $2\pi/k\to\pi/k$ issue}
Correspondingly, for each mode $b^{(m)}$, we divide $L$ evenly into $n_m=k_m L/\pi$ cells. 
For most lines of sight, 
%$m$, 
$n_m$ is  greater than $L/l$ since roughly the largest mode has a wavelength no larger than  $l$.
The length of the line of sight inside the galaxy will typically be of order  $\llg$, the characteristic scale of a large-scale magnetic field, except when observations are made edge-on and close to the galactic outer edge. 
Therefore, if we assume that  $L/l> 1$,  we have $n_m>1$. 
Large $n_m$ will allow more accurate  application of  the CLT.
%EB20 tweaked above

In  each separate cell of scale $\pi/k_m$, 
$b^{(m)}$ is nearly coherent in space with the same sign,  parallel or anti-parallel  to $d\bm s$. 
We can then replace $b^{(m)}$ by its root-mean-square value $b_m$ defined by
a MFE-appropriate average (section \ref{sec:2.3}), supplemented by a `+' sign if  parallel to $d\bm r$, and a `-' sign if  anti-parallel. 
Then equation (\ref{b_L})  becomes the sum of $m$  averages, each  being the mean of $n_m$ random variables $b^{(m)}_i$, taking a value $b_m$ or $-b_m$:
\beq
b_L=\frac{1}{L}\int d\bm s\cdot\bmb
=\frac{1}{L}\sum_m \int ds\ b^{(m)}
=\sum_m \frac{1}{n_m} \sum_{i=1}^{n_m} b^{(m)}_i.
\eeq
%In statistics, central limit theorem (CLT) says that, the sum of $n$ independent random variables with the same probability distribution with a mean $M$ and variance $V$, will be asymptotically normally distributed as $n\rightarrow\infty$ with a mean $M$ and variance $V/n$. 
%\remark{HZ edited below}
Although 
$b^{(m)}_i$ is likely to be correlated with
%\remark{HZ}
both  its spectral neighbor $b^{(m+1)}_i$ and spatial neighbor $b^{(m)}_{i+1}$ because  the turbulent fields are entangled locally in both configuration and Fourier space, 
we assume  that every $b^{(m)}_i$ varies independently and leave  generalizations for future work.

For  $n_m\gg1$  the scale separation is large and 
$\sum_i b^{(m)}_i/n_m$ is close to a normally distributed random variable with zero mean and variance $b_m^2/n_m$.  Then 
$b_L$ is the sum of $m$ independent normally distributed random variables and thus  a random variable itself, with variance \citep[][p. 256]{RSS1988}
\beq
\errordf=\sum_m \frac{b_m^2}{n_m}=\frac{1}{L}\sum_m \frac{\pi b_m^2}{k_m}.
\label{s^2}
\eeq
%where $(...)$ represents correlations between different modes, 
%Most literatures assume $\lra{b(\bmk)b(\bmk')}\sim\delta(\bmk+\bmk')$. 
%We adopt this relation and therefore these correlation terms vanish, leaving
%\beq
%\sigma^2=\frac{1}{L}\sum_m \frac{2\pi b_m^2}{k_m}.
%\eeq
The summation on the RHS in equation (\ref{s^2}) is the energy density-weighted average wavelength up to a constant. The  relation to energy density  is somewhat of a  coincidence arising because both energy and variance are related to $\lra{{b^{(m)}}^2}$. 

The variance is more  useful in its integral form.
Let ${\utilde M}(k)$ be the energy spectrum of the total magnetic field.
In general, $\utilde M (k)$ could vary in space, but  for line-of-sight measurements, the  energy spectrum averaged over the line of sight is a reasonable approximation.
The energy spectra of large- and small-scale fields are then  $|\uGl(\bmk)|^2 \utilde M(k)$ and $|1-\uGl(\bmk)|^2 \utilde M(k)$, respectively.
Hence $b_m^2$ is related to the energy spectrum through
%\remark{HZ edited below}
\beq
\frac{b_m^2}{8\pi}=|1-\uGl(\bmk_m)|^2{\utilde M}(k_m)dk_m,
\label{4.8}
\eeq
given that $\uGl$ is isotropic.
Using  this and the integral version of  equation (\ref{s^2}), we obtain
\beq
\errordf
=
\frac{8\pi^2}{L}\int_{0}^{k_\nu} dk\ \frac{|1-\uGl(\bmk)|^2{\utilde M}(k)}{k}=
\frac{8\pi^2}{k_{\text{int}}L}\int_{0}^{k_\nu} dk\ |1-\uGl(\bmk)|^2{\utilde M}(k),
\label{var}
\eeq
where $k_\nu=2\pi/l_\nu$ is the wavenumber of the dissipation scale, 
and
%\remark{HZ edited below}
we have defined
\beq
k_{\text{int}}\equiv
\frac
{\int_{0}^{k_\nu} dk\ |1-\uGl(\bmk)|^2{\utilde M}(k)}
{\int_{0}^{k_\nu} dk\ |1-\uGl(\bmk)|^2{\utilde M}(k)/k}
\eeq
to be the integral scale of fluctuations which depends weakly on $l$
 but roughly equals  $\pi/\lsm$, since $\lsm$ is the coherent scale and the wavelength corresponding to it will be $2\lsm$.

Equation (\ref{var}) reveals that  $\errordf$ is proportional to the total magnetic energy  in fluctuations, and the ratio between $\pi/k_{\text{int}}\simeq\lsm$ and the segment length $L$  through the source. 
%That is  the main result of this section.
%HZ4 very good,minor tweak
Equation (\ref{var}) is testable with simulations.
The ensemble associated with the standard deviation on its LHS could be realized by taking snapshots of the system at different times (which would  equate the time average to an ensemble average),
whereas the integral on the RHS is measurable  in Fourier space. 

%Both $k$s on the bottom and in $E(k)$ in the integrand stand for (the inverse of) the scale of variation of small-scale field amplitudes. 

To illustrate the use of equation (\ref{var}),
we assume  $\pi/k_{\text{int}}=\lsm$ and define 
\beq
q_l=\frac
{\int_0^{k_\nu} dk\ |1-\uGl|^2 \utilde M}
{\int_0^{k_\nu} dk\ \uGl^2 \utilde M}
\label{ql}
\eeq
as the proportionality between small and large scale magnetic energies.
The $q_l$ is independent of location along each line of sight but depends upon how we define large- and small-scale fields, through  $l$ and $\Gl$.
Hence equation (\ref{var}) yields
\beq
\errordf
=\frac{\lsm}{L}
\left(\frac{\int dk\ |1-\uGl|^2 \utilde M}{\int dk\ \uGl^2 \utilde M}\right)
\left(8\pi \int dk\ \uGl^2 \utilde M\right)
=\frac{\lsm}{L} q_l (\bar B^2)_L
\label{eqn_var'}
\eeq
where in the last equality, $(\bar B^2)_L/8\pi=\int dk\ \uGl^2 \utilde M$, the line-of-sight average of the large-scale field energy.
Note that $(\bar { B}^2)_L$  is distinct from $(\bar B_L)^2$ as the latter is the square of the line-of-sight average [see equation (\ref{eqn_los_average}) of the theoretically predicted mean field $\bar\bmB$].
%EB2 thinking reader may be confused a little about whether  ${\bar B}_L=  |\bar {\bf B}_L|$ or  $|\bar {\bf B}\cdot {\hat{ \bf r}}|_L$
%HZ2 i added reference to eqn. 4.2 above. 
%EB12 In the above equation and that below, wouldnt we have $\xi ~ \lsm/l$ as if 2pi/k_int =\lsm as per discussion below 4.10. That conclusion does  depend on the spectrum though and works if  we choose l above the scale where  the small scale spectrum peaks. Otherwise if spectrum is still rising toward scale l then it would  seemingly be l.
%EB12 shortened this section to reflect our revision above
%The insensitivity of  replacing smooth filtering by sharp cutoff
% in the second equality
%  was verified in \cite{ShapovalovVishniac2011} for the magnetic helicity spectra.
%EB12 tweaked above

To express $q_l$ in terms of $l$, we assume that $\lsm$ and $\llg$ are insensitive to $l$. 
We have checked that this is justified if, regardless of shape,  $\utilde M (k)$ has two peaks, one near $k=\klg=2\pi/\llg$ and  one near $k=\ksm=2\pi/\lsm$, and  is small near $k=k_l$.
%\remark{HZ edited below}
We also define $q\equiv\lra{b^2}/\lra{B}^2$ as the 
proportionality between the unfiltered small- and large-scale magnetic fields [= ratio of areas
under the two peaks of  $\utilde M(k)$].
%\remark{EGB:  Above we used $\langle \rangle$ for ensemble averages, so maybe we should make sure notation is not confusing}
Observations indicate that $q $ is on average somewhere between 3 and 4 \citep{Fletcher2010,VanEckBrownShukurovFletcher2015, Beck2016}; we adopt a fiducial value $q=4$.
Consequently, we have 
%\remark{EGB: definition of $q_l$ below uses absolute value where in Eq \ref{ql}  it just has $( )$,
%seems we should also use $()$ below }
%\remark{HZ: changed to absolute values everywhere}
\begin{align}
q_l
=&\frac
{\int dk\ |1-\uGl|^2 \utilde M}
{\int dk\ |\uGl|^2 \utilde M}\notag\\
\simeq&\frac
{|1-\uGl(\klg)|^2 \lra{B}^2+|1-\uGl(\ksm)|^2 \lra{b^2}}
{|\uGl (\klg)|^2 \lra{B}^2+|\uGl(\ksm)|^2  \lra{b^2}}\notag\\
=&\frac{|1-\uGl(\klg)|^2+|1-\uGl(\ksm)|^2 q}{|\uGl(\klg)|^2  +|\uGl(\ksm)|^2 q}.
\label{eqn_areaapprox}
\end{align}
Combining equations (\ref{eqn_var'}) and (\ref{eqn_areaapprox}) we have
\beq
\errordf
=\frac{\lsm}{L} q_l (\bar B^2)_L
=\frac{\lsm}{L}
\frac{|1-\uGl(\klg)|^2+|1-\uGl(\ksm)|^2 q}{|\uGl(\klg)|^2  +|\uGl(\ksm)|^2 q}
(\bar B^2)_L.
\label{var_result}
\eeq
%EB12 the paragraph below should be revised if we are ok with 4.12 in its new approximate form 
Equation (\ref{var_result}) highlights that the variance in  predicted RM 
is  the product of three factors. 
%First,  the ratio between the scale of average and the length of the line of sight. $l/L$.
First, the inverse of the number of eddy cells along the line of sight, $\lsm/L$.
Being linear in the length ratio, this can be significant even when  the  correction terms to the modified MFE equations are small.
The MFE corrections are of order $(l/\llg)^c$ [see equation (\ref{gamma_hat})], so for  small $l/\llg$ ratio or large $c$, the corrections could be small even if 
$\errordf$ is significant.
Second, $\errordf$ depends on  how energy is distributed between large and small-scale fields through  $q_l$. 
Since a larger $l$ implies  more modes are counted as  small scale fields ($k\lesssim 2\pi/l$), $q_l$ is a monotonic function of $l$.
Finally, equation (\ref{var_result}) shows that 
$\errordf$ is also proportional to the average magnetic energy density along the line of sight. 
%In disc galaxies, $(\bar B^2)_L/{(\bar B)}^{ 2}_L$  typically exceeds unity because azimuthal fields 
%dominate.
%EB2 the reader may think the above ratio exceeds unity by definition
%EB2 thinking out loud, we can discuss the above reasoning 
%HZ The turbulent scale is \lsm, and L>l>\lsm>l_v

Some complexities of the true error are not considered in equation (\ref{var_result}).
First, due to local inhomogeneities (spiral arms for example), cells for each mode along a single line of sight may not be statistically identical nor have the same total amplitude of  fluctuating magnetic energy as we have assumed.
%have identical  energy spectra
%  and steady states of turbulent fields, because transport coefficients, fluid motions and source fields from stellar activities vary from place to place in a galaxy. 
%Spatially fluctuating properties of cells bring more uncertainties in FR measurements, and in order to overcome it,
In equation (\ref{4.8}) we have used the line-of-sight averaged energy spectrum $\utilde M (k)$ as an approximation and ignored spatial variation of $\lra{b^2}$.
% leaving a more comprehensive formalism for future work.
Second, differential rotation makes  turbulent magnetic fields anisotropic in an eddy turnover time $\tau$ in the galactic mid-plane.  The azimuthal fluctuation  is amplified beyond the radial field such that $b_\phi\simeq b_r(1+q_r\Omega\tau)\simeq 2b_r$ with the $q_r\simeq 1$ for a flat rotation curve, and Rossby number $Ro=1/(\Omega \tau)\approx 1$ in spiral galaxies.
Therefore, the two components $b_r,b_\phi$  contribute unequally along different lines of sight, making FR measurements depend not only on $L$, but also on direction.

\section{Galactic dynamo and  precision   for different FR  viewing angles} 
\label{sec_application}
In this section  we  consider  specific cases  to elucidate the application of the calculations  of precision of mean field theories given by equations (\ref{errorint}) and (\ref{var_result}) in the context of  FR measurements.
We  calculate $\sigma^2=\errorint+\errordf$ in terms of mean fields when the measured galaxy is edge-on, face-on, and inclined. 
We  also consider the special case of measuring FR from within  our own Galaxy. 
We use a cylindrical coordinate system centered at the galactic center with coordinates $(r,\phi,z)$ %or $(\rho,\phi,z)$,
 and the $z$ axis coinciding with the galactic rotation axis.

\subsection{Galactic dynamo model}
\label{sec_dynamomodel}
We augment the simplified galactic dynamo model from section 4.5 of \cite{ZhouBlackman2017}
\footnote{Use of this model is intended to exemplify the method. Other models 
\citep[e.g.][]{Chamandy2016a} can also be used.},
where the `no-$z$' approximation \citep{SubramanianMestel1993,Moss1995,Phillips2001,SurShukurovSubramanian2007,Chamandy2014} is used.  The resulting $\bmB$ is $r$-dependent and cylindrically symmetric (i.e., azimuthally averaged).  We include  the correction terms of section \ref{sec_dynamo} employing a Gaussian kernel
%\remark{HZ added below}
(and thus $\hg=-\l^2 \nabla^2/8\pi^2$), which gives 
\beq
\hg=-\frac{1}{2}\frac{l^2}{4\pi^2}\partial_z^2 \to \frac{l^2}{32h^2},
\label{noz}
\eeq
%\remark{HZ edited below}
where derivatives in the radial direction are dropped assuming the disk is thin, $h/R\ll1$.
The last relation in equation (\ref{noz}) follows from the `no-$z$' approximation,
%HZ30
%EB30 good
$\partial_z^2 \to -(k_h/4)^2$
where $k_h=2\pi/h$
\citep{Phillips2001, SurShukurovSubramanian2007}.
%HZ7
To the helicity density evolution equation with  flux terms \citep{BrandenburgSubramanian2005, SubramanianBrandenburg2006,SurShukurovSubramanian2007} we also add the  correction terms resulting from violation of the Reynolds rules and obtain 
%\beq
%\partial_t\chi +\del\cdot\bm F=-2(1-\hg)(\emf\cdot\bar\bmB)-2\bar\bmB\hg\emf-2\nu_m\bar{\bm j\cdot\bmb}
%\eeq
\beq
\partial_t\alpham
=-\frac{2\beta}{\lsm^2}
\left[
\frac{(1-\hg)(\emf\cdot\bar\bmB)+\bar\bmB\hg\emf}{B_{eq}^2}
+\frac{\alpham}{R_\text{m}}
\right]
-\del\cdot(\alpham\bar\bmU)
-\beta_d \nabla^2\alpham.
\label{eqn_helicity}
\eeq
The last term of equation (\ref{eqn_helicity}) governs the diffusive flux and we adopt $\beta_d=\beta$.

%\remark{HZ edited below}
With the requirement that $l<h$, we find that the  $\hg$ correction terms  produce only small changes in the  dynamo model solutions   and we can omit them in the later discussion of the precision error.
%HZ30
%EB30 tweaked paragraph below, is it ok?
%HZ31 removed "compared to those without"
However, the smallness of the effect on the solutions  is a  feature of  our particular dynamo 
 model that is  exacerbated by  the  aforementioned `no-$z$' 
approximation.  To see this note that 
for our choice of  $l$, $\utilde\gamma(k_h)<1$ and  the magnitude of $\hg$ is always less than $1/16$.
The maximum value of $(\bar B)_L(r)$ when $l$ is increased from $0.1h$ to $0.9h$ from the solutions changes by just $\sim1\%$.
If instead we had  used the approximation that  $\partial_z^2 \sim -k_h^2$,  there would be a $\sim 40\%$ decrease in the maximum value of $(\bar B)_L(r)$ when $l$ is increased from $0.1h$ to $0.9h$ from the solutions with the correction terms.  
This highlights that  the correction terms are not necessarily small for every model. 
Moreover, in the absence of any  significant scale separation between large and small scale parts of the magnetic energy spectrum, the expansion of $\uGl(\bmk)$ in 
%$\hg$
Eqn. (\ref{gamma2.4}) would itself be invalid, and corrections to the MFE would be non-perturbative.
% must would have to be treated  as an operator without any approximations.

Numerically, \cite{ShapovalovVishniac2011}   found,  from the 
%uncorrected 
(uncorrected) evolution equation of small-scale helicity,  
that the resultant spectra of large-scale quantities are  
insensitive to different filtering methods, for reasonable spectra of relevant total quantities.
%EB20 tweaked above 

The steady state\footnote{
%	\remark{HZ added footnote here}
	Here we focus on a time-independent field (as a valid and simple solution to the dynamo model) to illustrate the idea of quantifying precisions of a mean field theory.
	In principle, similar calculations can be done at each instant time for a non-steady state (e.g., oscillatory) solution.
}, non-dimensionalized dynamo equations read
\begin{align}
0
=\partial_tB_r
=&-\frac{2}{\pi}R_\alpha(1+\alpham) B_\phi
-\left(R_U+\frac{\pi^2}{4}\right)B_r\label{rde1}\\
0
=\partial_tB_\phi
=&R_\omega B_r
-\left(R_U+\frac{\pi^2}{4}\right)B_\phi\label{rde2}\\
0
=\partial_t\alpham
=&-R_U\alpham-\frac{\beta_d}{\beta}\frac{\pi}{2}\alpha_m
-C\left[
(1+\alpham)(B_r^2+B_\phi^2)
\right.\notag\\
&\left.
+\frac{3}{8}\sqrt{\frac{-\pi(1+\alpha_m)R_\omega}{R_\alpha}}B_rB_\phi
+\frac{\alpha_m}{R_m}
\right],
\label{rde3}
\end{align}
where
\beq
R_\alpha=\frac{\alphak h}{\beta},\ 
R_U=\frac{|\bar\bmU|h}{\beta},\ 
R_\omega=-\frac{h^2 \Omega}{\beta},\ 
C=2\left(\frac{h}{\lsm}\right)^2
\label{eqn_dimensionlessparameters}
\eeq
are dimensionless parameters with a flat rotation curve $\Omega\propto 1/r$ adopted,
and magnetic fields are normalized by the equipartition field strength $B_{\text{eq}}=\sqrt{4\pi\rho_{\text f} u^2}$ with $\rho_{\text f}$ being the fluid density.
The $\alpha$-coefficients are normalized by $\alphak$.
The $r$-dependence of equation (\ref{eqn_dimensionlessparameters}) is described in detail in section 2.4 and equation (41) in \cite{ZhouBlackman2017}.
The approximation for the $\emf\cdot\bar{\bmB}$ term can be found in the appendix of \cite{SurShukurovSubramanian2007} or that of \cite{Chamandy2013}.

Analytical expressions of $\bar B_\phi$ and $\bar B_r$ are obtainable from equations (\ref{rde1}) to (\ref{rde3}).
The intrinsic error of $\bar\bmB(\bmx)$ is then given by equation (\ref{errorintbi}) in terms of $\sgma^2$, $\sgmb^2$ and $\sgmab$.
The first two are given in equation (\ref{sigma_a_and_b}), where as for $\sgmab$ we assume that fluctuations of $\alphak$ and $\beta$ are uncorrelated, and 
\beq
\sgmab
\simeq(\sgma^2 \sgmb^2)^{1/2}
=\sgmb^2 R_\alpha/h.
\label{alphabetaerror2}
\eeq
(For galaxies, $R_\alpha\simeq 1$.)
At a fixed location, $\bar\bmB$ is a function of $R_\alpha$, $R_U$ and $R_\omega$.
Therefore the partial derivatives with respect to $\alphak$ and $\beta$ can be evaluated using the chain rule,
\beq
\partial_{\alphak}=\frac{h}{\beta}\partial_{R_\alpha},\ 
\partial_\beta=-\frac{1}{\beta}\left(
R_\alpha \partial_{R_\alpha}+R_U \partial_{R_U}+R_\omega \partial_{R_\omega}
\right).
\label{chainrule}
\eeq
Combining equations (\ref{errorintbi}), (\ref{alphabetaerror2}) and (\ref{chainrule}), we have for the intrinsic error of $\bar B_i$,
\begin{align}
\errorintb
=\frac{1}{12(l/\lsm)^3}\left\{\right.&
(\partial_{R_\alpha}\bar B_i)^2
R_\alpha^2
+[\left(
R_\alpha \partial_{R_\alpha}+R_U \partial_{R_U}+R_\omega \partial_{R_\omega}
\right) \bar B_i]^2
\notag\\
&\left.
-2(R_\alpha\partial_{R_\alpha}\bar B_i)
[\left(
R_\alpha \partial_{R_\alpha}+R_U \partial_{R_U}+R_\omega \partial_{R_\omega}
\right)\bar B_i]
\right\},
\label{errorintbi2}
\end{align}
and that of $(\bar B)_L$ is given by substituting equation (\ref{errorintbi2}) into equation (\ref{errorint}), given the solutions of equations (\ref{rde1}) to (\ref{rde3}).

\subsection{Edge-on view}
We first consider a special case  representing  the measurement  of  FR  of a perfectly edge-on disc galaxy with radius $R=12\ \text{kpc}$ (see the schematic diagrams of  figure \ref{schematic_plots}). 
Note that the integration path segments  along the line of sight  within the galaxy form chords  with lengths
$L(\varpi)=2\sqrt{R^2-\varpi^2}$, where $\varpi$ is the distance from the galactic center to the closest
point on the chord.
From the  geometry of the configuration, the line of sight average is 
\beq
\bar B_L (\varpi)=\frac{2\varpi}{L(\varpi)}\int_0^{L/2} dy\ \frac{\bar B_\phi(r)}{r},
\eeq
and 
\beq
(\bar B^2)_L (\varpi)=\frac{2}{L(\varpi)}\int_0^{L/2}dy\ \bar B^2(r)
\eeq
where $r=\sqrt{\varpi^2+y^2}$ is the radial coordinate from the galactic center. 
Only ${\bar B}_\phi$ contributes to $(\bar B)_L$
%\remark{HZ edited below}
for the edge-on view because $\bar B_r$ is mirror-symmetric about the $x$-axis and its contributions from the $y>0$ and $y<0$ regions cancel each other.
The intrinsic error is given by
\beq
\errorint
=\frac{2\varpi^2}{L(\varpi)}\int_0^{L/2} dy\ \frac{\sigma^2_{\text{int},\bar B_\phi}}{r^2}.
\eeq
The imprecision associated with the observation is given by equation (\ref{var_result}) and is
\beq
\errordf
=\frac{2\lsm}{L^2}
\frac{(1-e^{-l^2/2\llg^2})^2+(1-e^{-l^2/2\lsm^2})^2 q}
{e^{-l^2/\llg^2}+e^{-l^2/\lsm^2} q}
\int_0^{L/2}dy\ \bar B^2(r),
\label{sigma_edge_on}
\eeq
where we take $\llg\simeq h=0.5\ \text{kpc}$, for   galactic disk semi-thickness $h$, and the variation scale of turbulent fields $\lsm\simeq 0.1\ \text{kpc}$ is assumed to be the same for   velocity and magnetic fields.
Here  $\llg\simeq h$ because $\partial_r\ll\partial_z$ in a thin disk and $h$ is the smallest
natural scale of variation for the mean field. 
Correspondingly we take an averaging  scale  $0.12\le l \le 0.48\ \text{kpc}$. 

\begin{figure}
%EB20  the legend on the left should probably be changed so that the sigmas are listed with the colored dots, not with Bbar_L
	\centerline{\includegraphics[width=0.5\columnwidth]{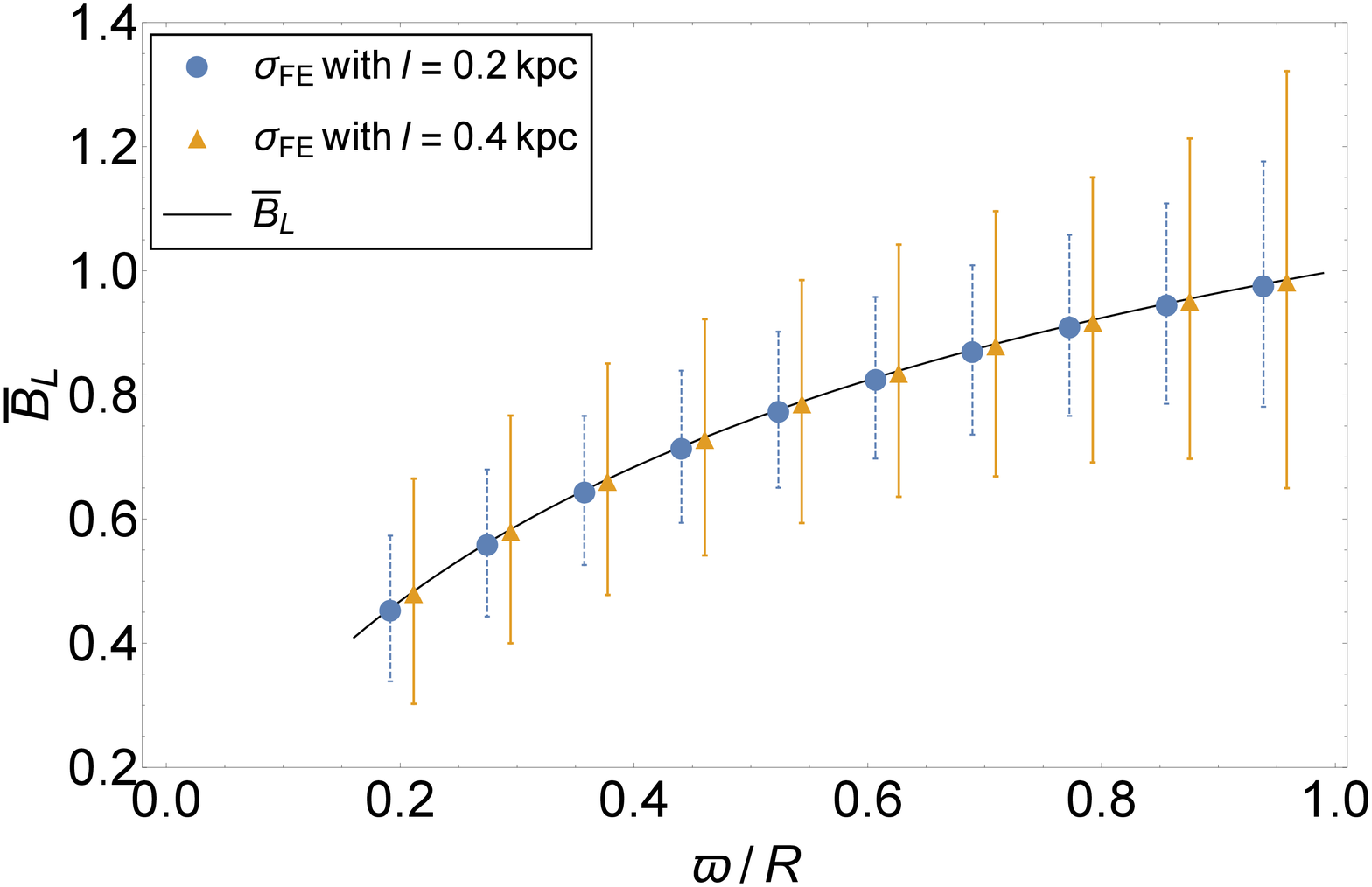}			
	\includegraphics[width=0.5\columnwidth]{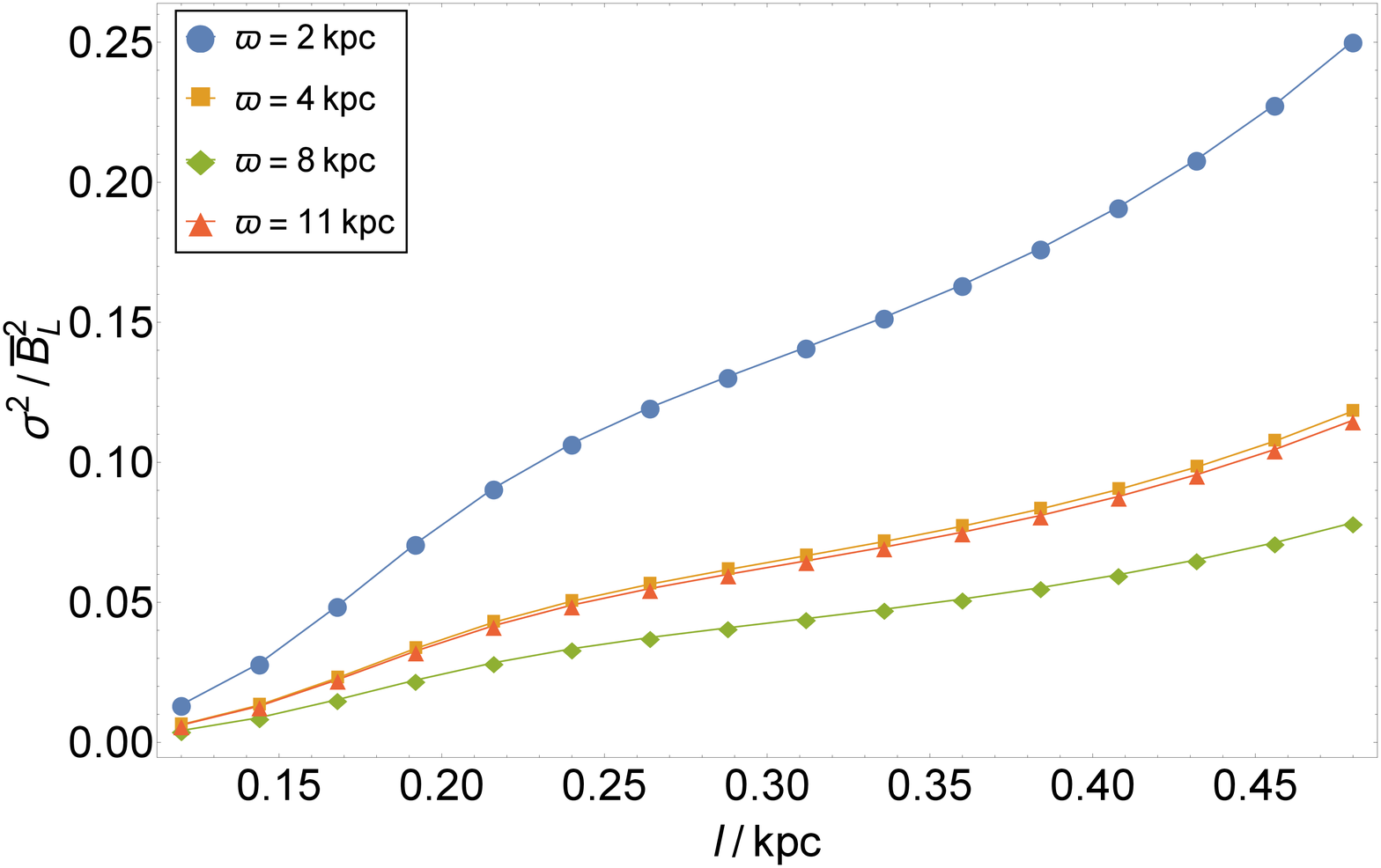}}
	\caption{
		The left panel shows theoretical predictions of the line-of-sight averaged magnetic field $\bar B_L$  with  the filtering error $\sigma_{\text{df}}$ shown as error bars in an edge-on view of a disc galaxy,
		%\remark{HZ edited below}
		assuming that the mean field has the form $\bar\bmB=B_0\hp$ with $B_0=1$.  Lengths are normalized by the galactic radius $R=12\ \text{kpc}$. Two sets of error bars are shown for different choices of $l$. Right panel shows the fractional error bar values at different radii as a function of the averaging scale $l$.	
	}
	\label{sigma_cst}
\end{figure}

\begin{figure}
	\centerline{\includegraphics[width=0.5\columnwidth]{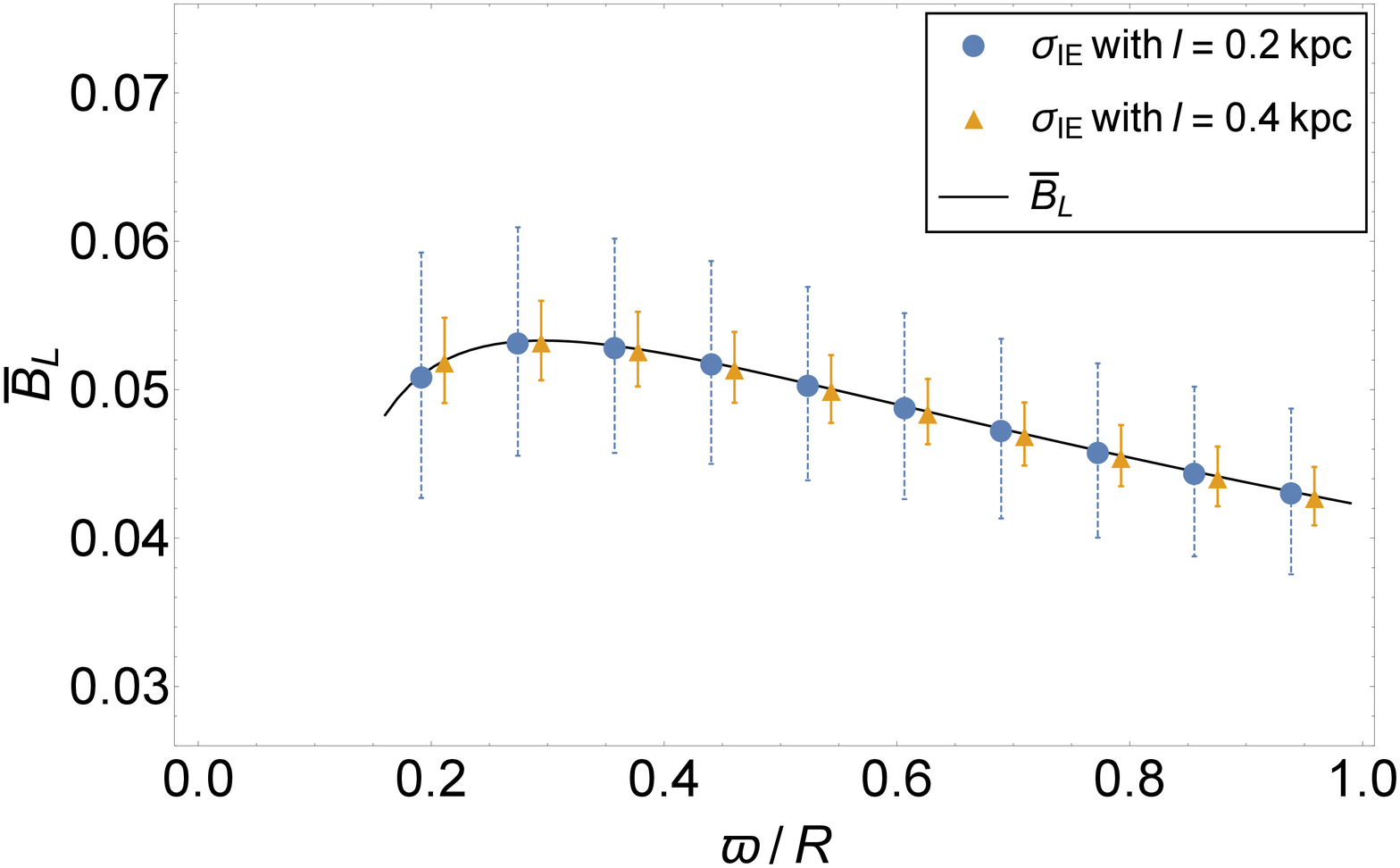}		
	\includegraphics[width=0.5\columnwidth]{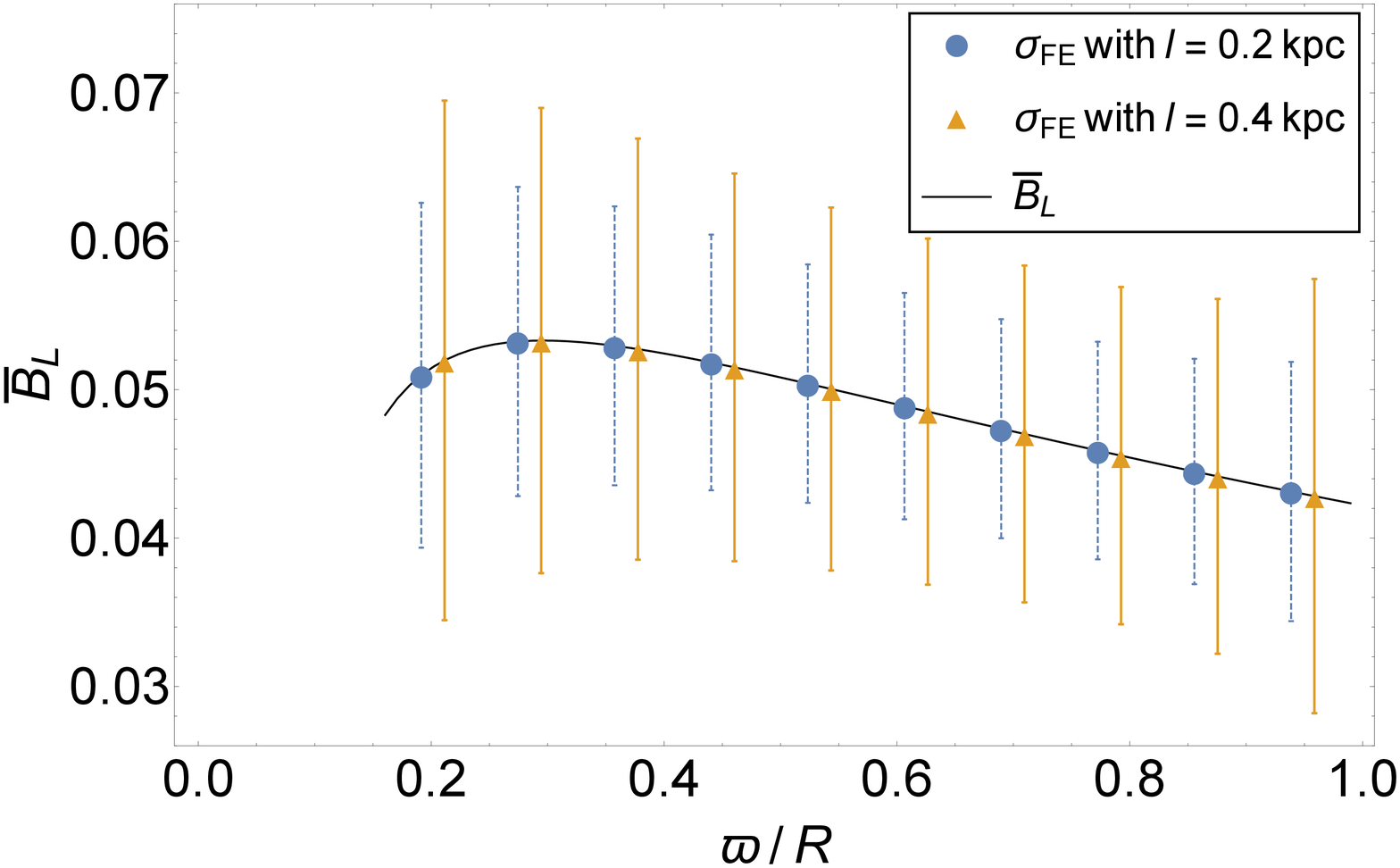}}
	\caption{
		Similar to the left panel of figure \ref{sigma_cst} but using  analytic  dynamo  solutions for $\bar\bmB$ from section 4.5 of \cite{ZhouBlackman2017}
		%\remark{HZ edited below}
		by solving equations (\ref{rde1}) to (\ref{rde3}).
		The left panel shows the intrinsic error and the right panel shows the filtering error.
	}
	\label{sigma_and_error}
\end{figure}

The predicted line-of-sight average of the magnetic field, together with the error bars are shown in figures \ref{sigma_cst} and \ref{sigma_and_error}, where two different profiles of $\bmB$ are separately considered: 
(i) in the left panel of figure \ref{sigma_cst} $\bar\bmB(\bmx)=B_0\hp$ where $B_0=1$ is a constant, and 
(ii) in figure \ref{sigma_and_error} the analytic solution of the  mean field dynamo model from section \ref{sec_dynamomodel}, normalized by the equipartition field strength $B_{\text{eq}}=\sqrt{4\pi\rho_{\text{f}} u^2}$.
%\remark{HZ edited below}
The dimensionless parameters we have used for the analytic solution are the same as those in \cite{ZhouBlackman2017}:
\beq
\Ra=R_{\alpha 0}/2,\ 
\RU=2R_{U0}/(r/\rsun)^2 F^{5/2},\ 
\Rw=2R_{\omega 0}/(r/\rsun)^2 F^3,\ 
C=4C_0/(r/\rsun)^2 F^3
\eeq
where quantities with subscripts $0$ are computed using
\begin{align}
&\tau_{\text{ed}}=10^{15}\ \text{s},\ 
u=10\ \text{km/s}^{-1},\ 
r\Omega=200\ \text{km/s}^{-1},\ \notag\\
&\lsm=0.1\ \text{kpc},\ 
h=0.5\ \text{kpc},\ 
U_0=1\ \text{km/s}^{-1},
\end{align}
which yields
\beq
R_{\alpha 0}=1,\ 
R_{U0}=0.3,\ 
R_{\omega 0}=-15,
\eeq
and we use 
$R_{\text{m}}=10^5$.
Above $\rsun\equiv 8\ \text{kpc}$ is the location of the Sun, and the function $F$ determines the $r$-dependence of the dimensionless parameters and is described in detail in the appendix in \cite{ZhouBlackman2017}.

The line-of-sight averages of the mean magnetic fields are shown as black solid curves, along with different types of error bars $=\pm \sigma$ about the mean computed from equations (\ref{errorint}) and (\ref{sigma_edge_on}) for the cases associated with two different choices of the scale of average, $l$. 
The blue dashed lines with circular markers give error bars with $l=0.2\ \text{kpc}$, and the yellow solid lines with triangular markers give those with $l=0.4\ \text{kpc}$.
In the constant magnetic field case, the intrinsic error does not exist because here $\bar\bmB$ is presumed, rather than derived from MFE equations.

%Recalling that the mean field corresponds to  roughly  the sum of modes with $k\leq 2\pi/l$, the fact that a smaller $l$ yields smaller error bars is a consequence of including more total number of small-scale modes into what comprises  the mean fields.
%In our formalism, the spectral space from the largest scale to $l$ includes more modes as $l$ decreases increasing the effective sample size for applicability of the central limit theorem, and lowering the variance.
%All else being equal, small $l$ therefore makes more accurate predictions for a  MFT. 
%Note that the minimal $l$ satisfies $l>\lsm$ for a well-defined statistical theory, thus naturally constraining the precision of the MFT to have a lower bound.
%For inhomogeneous or anisotropic turbulence, $\lsm=\lsm(\bmx)$ is a function of spatial coordinates by which the scale of average $l=l(\bmx)$ is bounded below.

Different choices of $l$ conspicuously show different levels of precision in the predictions for measurements, as evidenced by a  comparison of  the blue vs. yellow  IE bars  in the $r$-dependent model (left panel in figure \ref{sigma_and_error}). 
%are such that the  variation of the field is now submerged beneath the level of precision that the error bars of the theory.
Variations in a data curve beneath the level of the error bars cannot be  deemed a 
disagreement with the MFE theory.  That is,  whether uncorrelated or weakly correlated 
%EB21 added uncorrelated or weakly  -- then some extra info may be accessed perhaps.
deviations with amplitudes below the error bars
are systematic \citep{Chamandy2016b} or stochastic  is beyond the resolution of the theory.

Comparing the  panels of figure \ref{sigma_and_error}  highlights  competing dependences of  
$\errorint$ and $\errordf$ on $l$, as discussed in section \ref{sec_twoerrors}:
$\errorint$ grows with  $l$ but $\errordf$ decreases with  $l$.
Assuming  $\errorint$ and $\errordf$ are independent and uncorrelated, adding them in quadrature gives the total  uncertainty $\sigma^2$.

\begin{figure}
	\centerline{\includegraphics[width=\columnwidth]{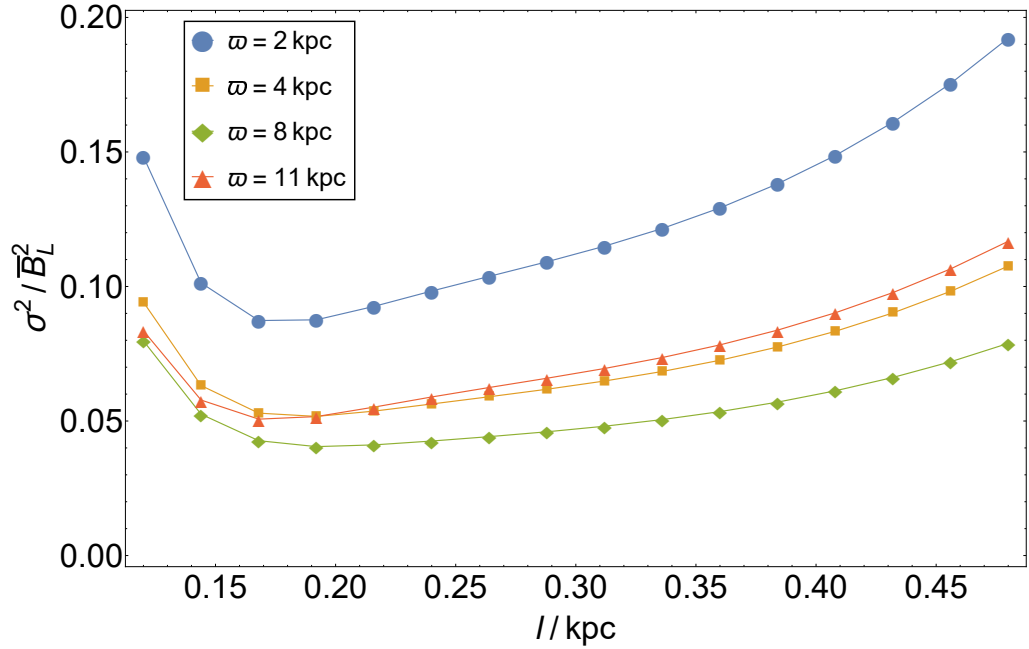}}
	\caption{
		The total relative error using the analytic dynamo solutions at different radii as a function of averaging scale $l$.
		An optimal scale arises at 0.15-0.20 kpc which minimizes the relative error, and therefore provides the best precision of theoretical predictions.
	}
	\label{fig_dynamo_rel}
\end{figure}

In the right panel of figure \ref{sigma_cst} and in figure \ref{fig_dynamo_rel}, we show the relative total errors, $\sigma^2/(\bar B)_L^2$, as a function of $0.12\ \text{kpc}\leq l \leq 0.48\ \text{kpc}$ at different galactic radii.
For figure \ref{sigma_cst} there is  only one uncertainty, namely  $\errordf$ which is a monotonic function of $l$
for all  radii shown.
%We  see 
%from the right panel of figure \ref{sigma_cst} that the
 %relative error is monotonic 
More interesting case is figure \ref{fig_dynamo_rel} where both $\errordf$ and $\errorint$ are competitive. There is an optimal scale of average, located at $0.15\le r/R\le 0.20$ for all four 
chosen radii, that minimizes the total error, and thus maximizes  the 
precision of comparing  theory  and observation.
In general. the existence and location of such a `sweet spot' depends on the solution to a given dynamo model, and the observational method used.

\subsection{Face-on view}
A complementary extreme to the edge-on case  is a  face-on  view. 
Here every line of sight is perpendicular to the galactic disk, taken along the $z$ direction. 
In this orientation,  $B_\phi$ and $B_r$ do not contribute to $(\bar B)_L$, and for a weak ${\bar B}_z$, the dominant non-vanishing RM would come from small-scale fluctuations. 
If we assume quasi-equipartition between the total mean and fluctuating small scale magnetic energies, the FR measurements still predict a a precision error about which the mean field is indeterminate.
%EB2 this would be somewhat challenged by observers sayi
% arge-scale fields fom stochastic RMs.

Taking $L=2h$, the thickness of the galactic disk, and noting that $\bar\bmB$ is solely a function of $r$ in equation (\ref{var_result}), we have
\beq
\errordf(r)
=\frac{\lsm}{2h}
\frac{|1-\uGl(\klg)|^2+|1-\uGl(\ksm)|^2 q}{|\uGl(\klg)|^2  +|\uGl(\ksm)|^2 q}
(\bar B_\phi^2+\bar B_r^2)_L.
\label{sigma_faceon}
\eeq
%EB2 added bar over B_phi
%where again, we take $q_=4$, and the range corresponds to the range of  $l=[0.1,0.4]\ \text{kpc}$.

\begin{figure}
	\centerline{\includegraphics[width=\columnwidth]{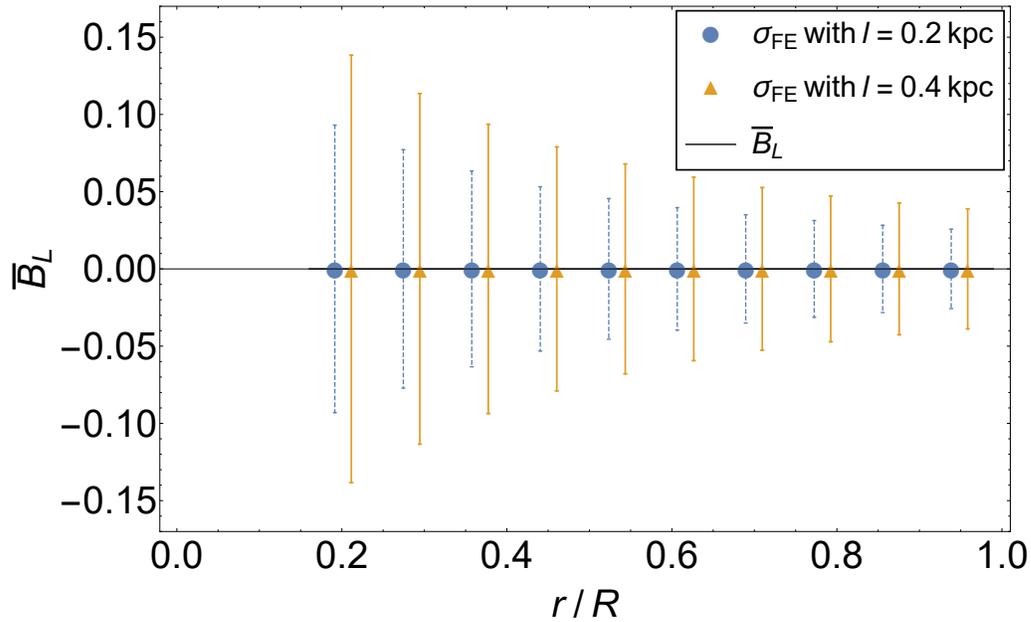}		
	}
	\caption{
		Similar to figure \ref{sigma_and_error} but for  a face-on view of a disc galaxy, using the analytic dynamo solution.
	}
	\label{fig_face_on}
\end{figure}

Figure \ref{fig_face_on} shows $\bar B_L$ as a function of the galactic radial coordinate $r$ (normalized by the galactic radius $R$) from a face-on view of the same $r$-dependent dynamo model used
in the last subsection \citep{ZhouBlackman2017}.
The predicted RM is now zero and its filtering error
 is given in blue dashed lines with circular markers for $l=0.2\ \text{kpc}$, and in yellow solid lines with triangular markers for $l=0.4\ \text{kpc}$.
 These  emerge purely from stochastic fluctuations.
 The intrinsic error is zero because $\bar B_z=0$ everywhere.

% and could  lead to a misperception that 
%EB20 it occurs to me , that if all data points say, were to lie above zero, then there is extra info in their mutual correlation that we are not taking into account"

\subsection{Views at intermediate inclinations}
The formulation becomes a bit more complicated when the line of sight is at an intermediate inclination.
%neither face on nor edge on
We  adopt Cartesian coordinates in this subsection, where the $z$-axis coincides with the galactic rotation axis, $x-y$ plane coincides with the galactic mid-plane,  the $y$-axis is parallel to the line of sight.
% lies in the plane 
%\remark{HZ edited below: EGB tweaked above}
 %containing the $z$-axis and parallel to the line of sight, and the 
Figure \ref{schematic_plots} shows a schematic plot.
Let the angle between the $z$ axis and the line of sight be $\theta$, and $0<\theta<\pi/2$.
The line-of-sight averages depend on the location of the intersection point of the line of sight and the galactic mid-plane, $(x,y)$, and are given by
\beq
(\bar B)_L(x,y)
=\frac{\sin\theta}{2h\sqrt{x^2+y^2}}\int_{-h}^h dz\ 
\left[
x \bar B_\phi(\rho)+y\bar B_r(\rho)
\right],
\eeq
and
\beq
(\bar B^2)_L(x,y)=\frac{1}{2h}\int_{-h}^h dz\ \bar B^2(\rho),
\label{5.7}
\eeq
where $\rho=\sqrt{x^2+(z\tan\theta+y)^2}$.
We  include only the region $\{(x,y)|\rho\leq R\}$.
%EB5 might be good to state that this is evident via the choice of integration bounds?
%
equation (\ref{5.7}) can then be used in equation (\ref{var_result}) to compute the precision error associated with FR measures, and the intrinsic error is given by
\beq
\errorint(x,y)
=\frac{\sin^2\theta}{2h(x^2+y^2)}\int_{-h}^h dz\ 
\left[
x^2 \sigma^2_{\text{int},\bar B_\phi}(\rho)
+y^2\sigma^2_{\text{int},\bar B_r}(\rho)
\right]
\eeq
which can be determined once  the intrinsic error of $\bar\bmB$ is calculated.

\subsection{View  from within  our Galaxy}\label{sec_measure_in_galaxy}
\begin{figure}
	\centerline{
		\includegraphics[width=\columnwidth]{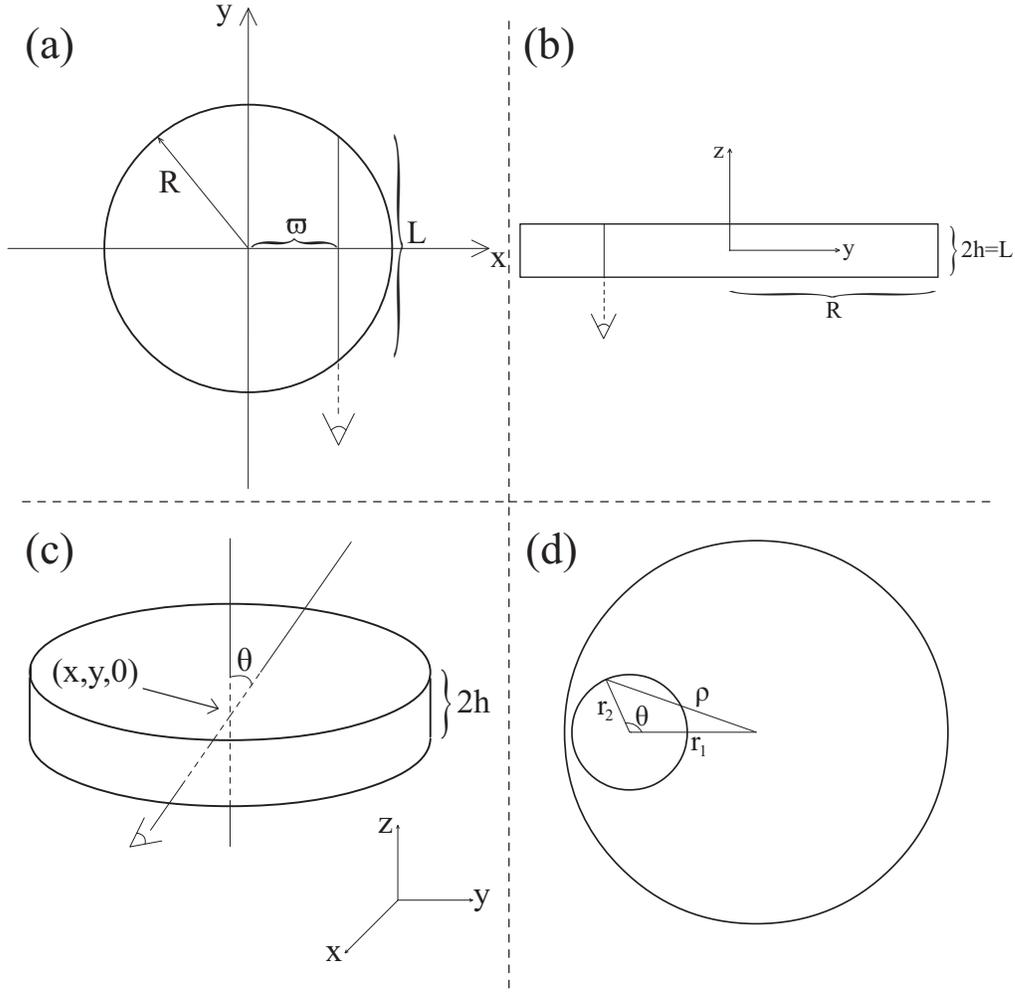}}
	\caption{
		Schematic diagrams of
%		\remark{HZ added below}
		line-of-sight averages for calculating the precision of RM in an
		(a) edge-on view, 
		(b) face-on view,
		(c) inclined view, and
		(d) inside the galaxy
		with $R$ being the galactic radius, $L$ the chord length along the line of sight, and $h$ the semi-thickness of the galactic disk.
		%\remark{HZ edited below}
		$\hat{\bm\rho}$ is the radial direction of the disk.
	}
\label{schematic_plots}
\end{figure}

\begin{figure}
	\centerline{\includegraphics[width=0.5\columnwidth]{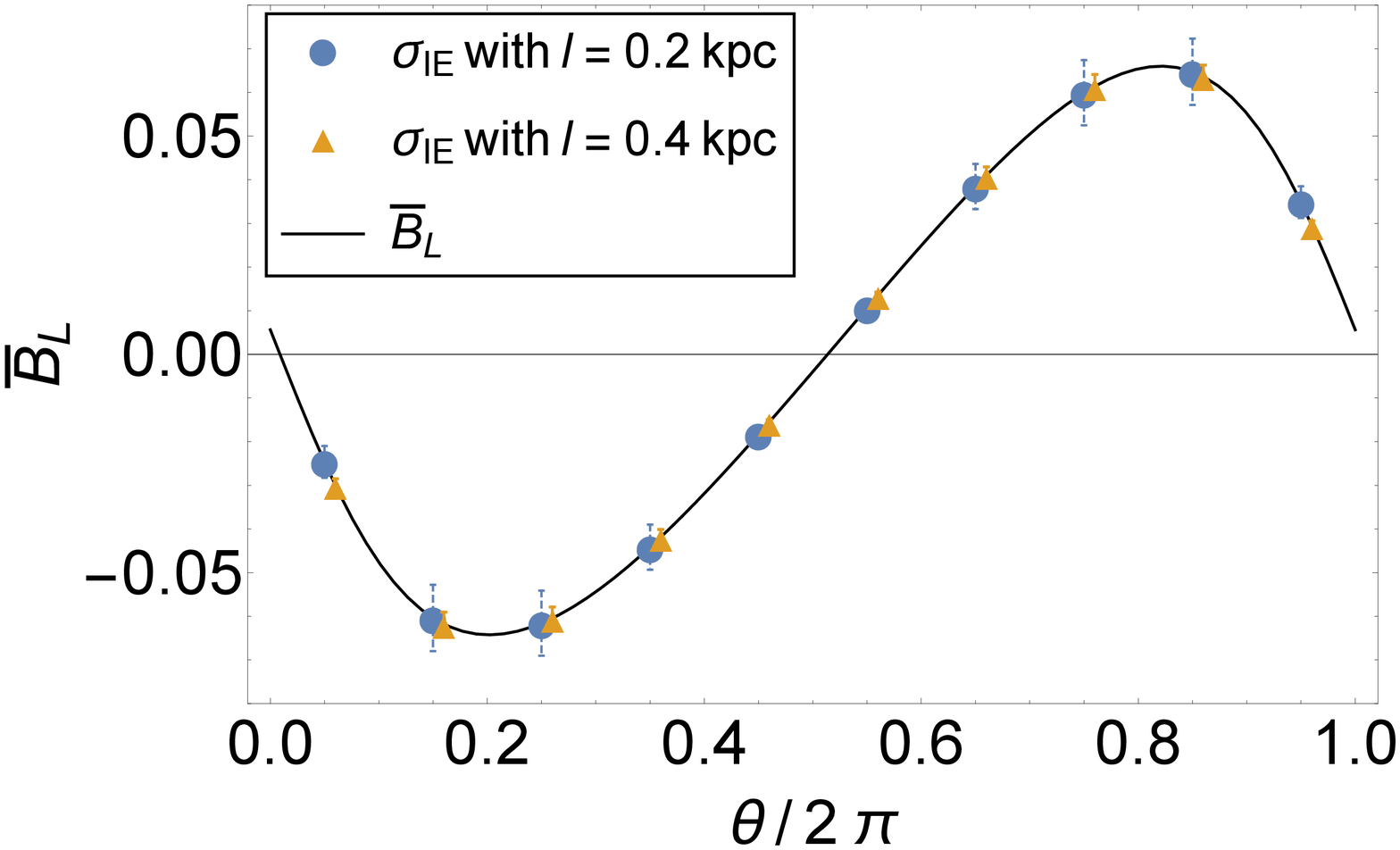}
		\includegraphics[width=0.5\columnwidth]{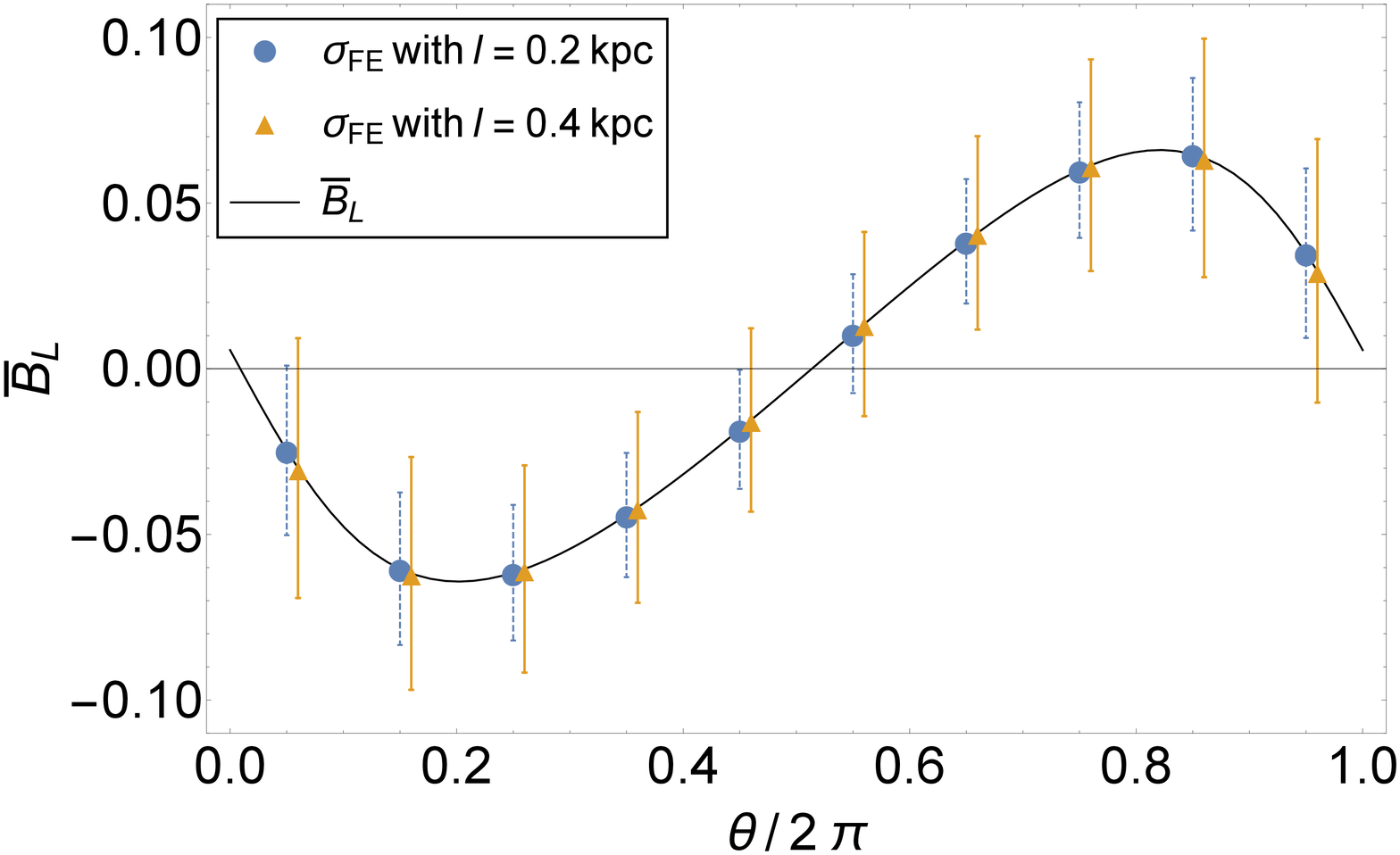}
	}
	\caption{
		Line-of-sight predictions and error bars of pulsar rotation measures for our  view from within our Galaxy based on the analytically solvable dynamo model,
		%\remark{HZ edited below}
		equations (\ref{rde1}) to (\ref{rde3}), taken from section 4.5 in \cite{ZhouBlackman2017}.
		Left and right panels show error bars corresponding to the intrinsic error and filtering error, respectively.
	}
	\label{fig_in}
\end{figure}

\begin{figure}
	\centerline{\includegraphics[width=\columnwidth]{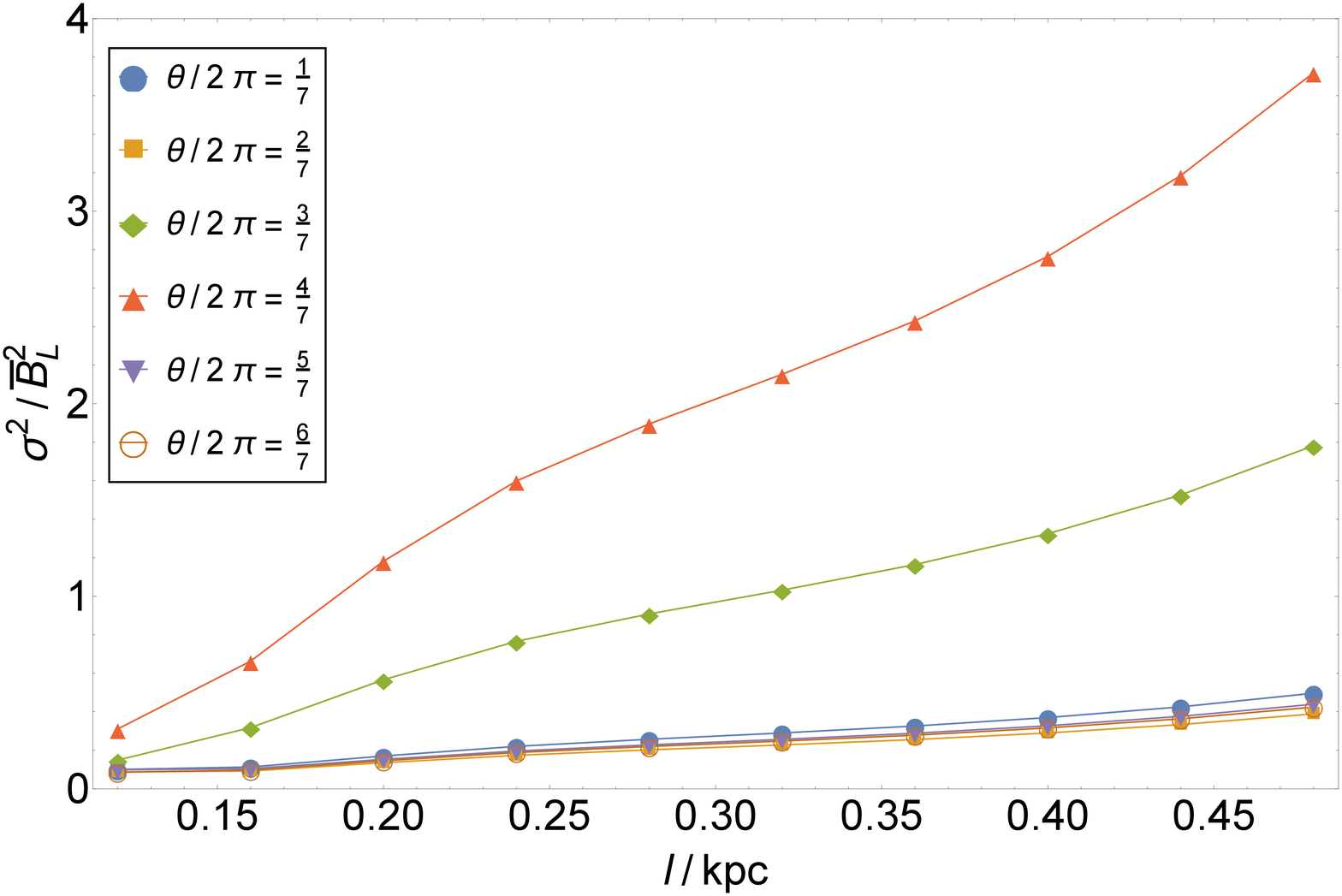}
	}
	\caption{
		The total relative error for pulsar RMs at different azimuthal angle (centered at the earth) as a function of averaging scale $l$.
		Filtering error dominates as a result of short length of the line of sight.
	}
	\label{fig_in_rel}
\end{figure}

Finally, we  discuss pulsar rotation measures as measured from inside our  Galaxy. 
For simplicity, we omit the $z$-dependence and assume that both the observer and pulsars are in the Galactic mid-plane.
A schematic plot is shown in figure \ref{schematic_plots}.
The distance of the observer to the Galactic center is denoted by $r_1$, and for this simple example,
we assume pulsars  to have a fixed distance $L=r_2<r_1$ from the observer and lie in the Galactic mid-plane.
We use $r_1=8\text{ kpc}$ and $r_2=3\text{ kpc}$ for typical values in calculations.
The line-of-sight average of magnetic fields is also a function of $\theta$, the azimuthal angle for a polar coordinate system centered at the earth which denotes the positions of pulsars, and $\theta=0$ points to the galactic center.
%EB20 state the definition of $\theta$ (centered around us not the galactic center and where $\theta=0 is$)
%\remark{HZ edited below}
The line-of-sight average of the mean field is then
\beq
(\bar B)_L(\theta)
=-\frac{r_1 \sin\theta}{r_2} \int_0^{r_2} dr\ \frac{\bar B_\phi(\rho)}{\rho}
+\frac{1}{r_2} \int_0^{r_2} dr\ \frac{r-r_1\cos\theta}{\rho}\bar B_r(\rho),
\eeq
where $\rho^2=r_1^2-2r_1 r \cos\theta+r^2$ is the radial coordinate in the galactocentric coordinate system (see figure \ref{schematic_plots}).
The line-of-sight averaged ${\overline B}^2$
%\remark{added overline to $B$}
is given by
\beq
(\bar B^2)_L(\theta)=\frac{1}{r_2}\int_0^{r_2} dr\ \bar B^2(\rho).
\eeq
The intrinsic error is given by
\beq
\errorint(\theta)
=\frac{r_1^2 \sin^2\theta}{r_2}\int_0^{r_2}dr\ \frac{\sigma^2_{\text{IE},\bar B_\phi}}{\rho^2}
+\frac{1}{r_2}\int_0^{r_2}dr\ 
\left(
\frac{r-r_1\cos\theta}{\rho}
\right)^2
\sigma^2_{\text{IE},\bar B_r}.
\eeq
The resultant curve is shown in figure \ref{fig_in} in the same plot style as those in the previous subsections.
%EB5 should the equation for (\bar B^2)_L be shown?
In this case, stochastic fluctuations  introduce only small $\errorint$ and 
moderate $\errordf$, the latter being dominant  
%$(\bar B)_L$, 
because the line-of-sight average yields a large $(\bar B)_L$ and the number of eddy cells along the line of sight is small as a consequence of small $L$. Thus $\errordf$ dominates the total uncertainty $\sigma^2=\errorint+\errordf$, and therefore in figure \ref{fig_in_rel} which again shows relative errors at different directions of observation as a function of $l$, most curves are monotonic and reach their minima when $l\rightarrow \lsm$.
Since $l$ is physically constrained in the region $[\lsm,\llg]$ (otherwise the statistical prescriptions of $\alpha$ and $\beta$ break down), this implies that $l\simeq \lsm$ is the optimal choice of average scale in this case.

It cannot be excluded that for different parameters, e.g. if  $q\equiv\lra{b^2}/\lra{B}^2$ were to exceed some critical value, 
the errors  might dominate mean field variations  making it difficult to statistically identify mean field reversals.
%EB20 tweaked above--ok?
%HZ20 yes

\section{Conclusions}\label{sec_conclusion}
\subsection{Summary}
For  large scale separation between mean fields and fluctuations, 
%HZ6 specified `ensemble average' below
ensemble and spatial averages are approximately equivalent, but this is not guaranteed in many astrophysical circumstances
where mesoscale fluctuations are present.
With this motivation, we  formally derived correction terms to MFE for spatial averaging that
result from a  finite  scale separation.
% and mean fields as is commomly
% so  we have derived correction terms to MFE
%raditional MFE of mesoscale fluctuations are identified.
%The previously commonly used ensemble average does not suit for our investigation, because the partition function associated with the ensemble of turbulence fields is unclear so far, and the practical utilization of ensemble average calls for assuming fluctuation scale greatly exceeds the scale of mean fields variations and equating spatial or temporal averages to ensemble average.
In addition, we have  quantified two types of MFE precision errors: 
%For our calculations of MFE precision, we showed how to compute two types of errors.
(i) the intrinsic error $\errorint$, which can be derived by differentiating the solution of the mean field equations with respect to its input parameters and propagating the uncertainty of each parameter to the mean field; and 
 %The result  is proportional to the number of eddy cells averaged $(l/\lsm)^3$ where $\lsm$ is the eddy scale, and 
%The deviation of mean field predictions for observed quantities compared to what is actually
%measured incurs 
(ii) the filtering error  $\errordf$ that results 
%\remark{EGB3: tweaked below}
because the prediction  from mean field theory is  filtered differently from the observations.
Specifically we considered the case where the predicted value is filtered using  the kernel for the mean field and then again by the measurement kernel --
whereas the observations    only singly filter the full field through the measurement kernel.

%To compute our averag this difficulty, an advantageous and also practical way for both theoretical and simulation researches
We derived the MFE corrections and precision errors
%using  is to define mean fields through
using  convolutions of the full field and  kernels, which introduce a prescribed averaging scale $l$.
To realistically  depict large-scale fields, the kernels must be chosen to be  local in both configuration and Fourier space, and monotonically decreasing in Fourier space.
We expanded the MFE equations in the ratio  $l/\llg$, where  $\llg$ is the dominant scale of variation of the mean field.
The  zeroth-order equations have the same form as those from  an ensemble average, but new first-order corrections of order $(l/\llg)^c$ arise due to a  violation of Reynolds rules, where $c>0$ depends on the   kernel.
Our approach allows for moderate scale separations.

% \cite{Rheinhardt2012}  assumed  a planar average for the magnetic field and from simulations, semi-empirically extracted the kernel relating the EMF to the mean magnetic field.
% In our case we start from first principles to focus on the  kernel of averaging the the mean
% field itself, and the correction terms from this kernel that mesoscale fluctuations  produce.
%Our approach can also be used to correct higher order terms. We also discussed the different role of the turbulent closure  in their approach and ours.

% is represented by the difference%
%between a singly averaged quantity and a doubly averaged quantity where the doubly averaged
%quantity has two types of averages.
 To exemplify the calculation of 
 %HZ30
 %these errors
 the precision errors, we considered  contributions to  (uniform density) galactic  Faraday rotation measures  from mesoscale fluctuations where the mean field filter  is a local spatial average and  the measurement kernel is  a line-of-sight average.
%With the assumption of quasi-equipartition between the energies of large- and small-scale fields, 
%The variation of RM can then be  expressed as the sum of the intrinsic part and the observational part, the latter being the product of three factors:
%(i) $\lsm/L$ where $L$ is roughly the size of the region with magnetic fields,
%(ii) $q_l$ which is the proportionality between the energies of large- and small-scale magnetic fields, and 
%(iii) $8\pi$ times the average energy density of large-scale magnetic fields along $L$.
We  applied the formalism 
%in section \ref{sec_application}
 to  different viewing angles of a disc galaxy and find that the precision error of MFE can be 
large even when the corrections to the  MFE equations themselves are small. This   highlights the necessity of quantifying this precision of mean field theories to avoid misconstruing stochastic from systematic deviations between theory and observations.  The error   quantifies the predictive resolution of the theory.

Since $\errorint$ decreases with $l$ while  $\errordf$ increases  with $l$, 
the sum of the two errors may be non-monotonic over the physically allowed range of $l$,
in turn allowing determination of optimal scale of $l$ that maximizes the precision of the theory.
For example,  we identified the optimal averaging scale for FR that minimizes the error  to be  about 0.17 kpc in our dynamo model for   edge-on galactic viewing.
%However, for  pulsar rotation measures discussed in section \ref{sec_measure_in_galaxy}, the smallest possible $l$ yields the best result.

%HZ6
%A related issue in statistical mechanics is the calculation of correction terms to mean field equations and uncertainties of mean quantities of an ensemble averaged system when the ensemble contains only limited number of elements.
%However, such calculations seem impossible without the knowledge of the distribution function or partition function.

%We have also identified in a toy model the uncertainties associated with pulsar rotation measures.
%Although the uncertainties are moderate compared to those with edge-on or face-on RMs, there are potential possibilities where standard deviations become large and thus could affect, for example, magnetic field reversal mapping.
We also  showed how our study    differs  from that of
 \cite{Rheinhardt2012} who were also motivated  to
address corrections to MFE equations for modest spatial scale separation.
Our focus is on the influence of the kernel  that enters the averaging of fields themselves whereas their focus was on the semi-empirically determined kernel relating the EMF to the mean magnetic magnetic field when the latter was defined through a planar average.
\subsection{Further work}
%EB10  tweaked this section
Our formalism can be  tested   and developed further.
First, using DNS for  a system that exhibits a statistically steady large-scale dynamo for a specific choice of  kernel average,  the saturated state from simulations could  be sampled at different times and an ensemble constructed. The mean field precision error   can then be measured and compared to 
our predictions. 
%Second, comparison and testing with DNS 
%EB5 lets chat about what specifically would be desirable here
%b) The assumptions of quasi-equipartition between mean fields and fluctuations and a power-law energy spectrum could be relaxed to seek for a more realistic model.
%c) 
Second, the MFE precision calculations that we exemplified for  FR  
could be generalized for more realistic numerical dynamo models, for  comparison to observations.
Generalization of the form of the magnetic spectra,  allowance for spatial inhomogeneities, 
or calculation of still  higher order corrections  to MFE equations are also  possible.
%HZ30
%EB30 tweaked below
Third, there remains analytical and numerical work 
to % rigorous treatment of  $\hg$ as an operator rather than making order-of-magnitude arguments;
%at even large $l$ 
study dynamo models in which the linear order corrections to the MFE equations are not
as small as those in the example models we considered with the `no-$z$' formalism.
For systems in which there is very little scale separation between large and small scale  energy spectral peaks, going  beyond our perturbative treatment of  Reynolds rules 
violations  would be necessary. 
The  resulting generalized MFE equations in this non-perturbative regime, with correction
terms that involve the full unexpanded kernel,  could be solved numerically. 
%This would be important when there is %regime where $\utilde\gamma(\klg)\gtrsim1$, 

More broadly, analogous computations of MFE precision  are warranted   for comparing theory and
observations for
%\remark{HZ edited below}
observables other than RMs such as polarized synchrotron emission in galaxies, or  
spectral fluxes in turbulent accretion disks. For the latter, the standard 
axisymmetric theory in common use is  also  an example of  a mean field theory which is a limiting case of MFE and has a  finite precision that has not yet been fully quantified \citep{Blackman2010}.

\noindent {\bf Acknowledgments: }
We are  grateful to referee Mathias Rheinhardt for providing numerous thoroughly perceptive and detailed comments  that helped us to very significantly improve the manuscript.  We acknowledge support from  grants NSF-AST-15156489 and  HST-AR-13916 and the Laboratory for Laser Energetics at U. Rochester. EB also acknowledges the Kavli Institute for Theoretical Physics (KITP) USCB and associated support from grant  NSF  PHY-1125915.

%====================Figures=====================

%EB20 i dont think we need the table below. its not referred to.
%\begin{table}
%	\begin{center}
%		\def~{\hphantom{0}}
%		\begin{tabular}{|c|c|c|c|c|c|c|c|c|c|c|}
%$\theta/2\pi$ & 0 & 0.1 & 0.2 & 0.3 & 0.4 & 0.5 & 0.6 & 0.7 & 0.8 & 0.9 \\
%$\errorint/10^{-3}$ & 0.19 & 3.58 & 6.50 & 7.90 & 6.25 & 0.26 & 6.25 & 7.90 & 6.50 & 3.58
%		\end{tabular}
%		\caption{
%			Inside. Intrinsic error. 
%		}
%		\label{tab_errorint}
%	\end{center}
%\end{table}
%============================================================
\appendix
\section{On the validity of equation (\ref{fml2'})}
\label{appx1}
In deriving equation (\ref{fml2'}), an approximation of $\bar{a \bar B}$, we have only considered the convolution of $\utilde a(\bmk)$ and $\utilde {\bar B}(\bmk')$ assuming 
that in this combination, only  small $k=|\bmk|$ and $k'=|\bmk'|$ contribute.
%As noted by one of our referees, 
 There are also contributions from other combinations of $k$ and $k'$.
In this appendix we discuss and quantify the validity of equation (\ref{fml2'}), 
 and show that it depends primarily on the scale separation $l_s/L$.
Specifically, we show  that for a Gaussian kernel with $k_l=5$, equation (\ref{fml2'}) is a good approximation when $\ksm/\klg\gtrsim 20$, assuming both $\utilde A(k)$ and $\utilde B(k)$ are double-peaked, and $\ksm$ and $\klg$ are the characteristic wave numbers of small and large scales, respectively. In this respect, the approximation we use 
improves the standard theory by relaxing the assumption of infinite scale
separation, but is  not valid for arbitrarily small separation.
% We note that for a galaxy,typically  $10 \ge k_s/k_L \le 20$.
%For example in these plots I showed to you today, for large $\ksm/\klg$ almost all the contributions come from (small $k$, small $K$), but in the referee's handwritten plot it seems like it is (medium $k$, medium $K$) which matters.

\begin{figure}
	\includegraphics[width=\columnwidth]{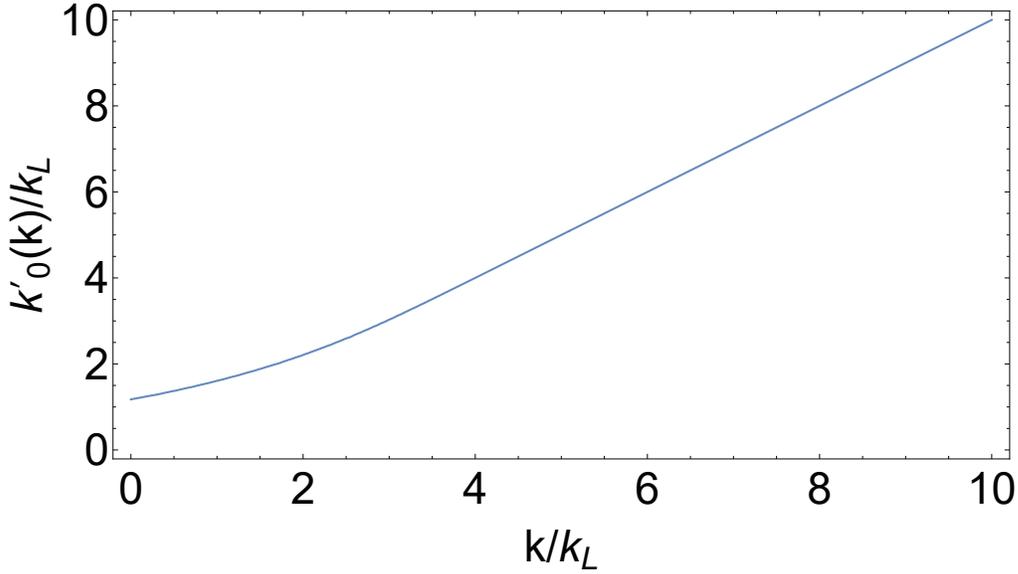}
	\caption{
		$k'_0(k)$ for a Gaussian kernel $e^{-k^2/2k_l^2}$ with $k_l=1$.
	}
	\label{fig_k0}
\end{figure}

For simplicity, we focus on one-dimensional cases here.
Which parts in the spectra of $a$ and $\bar B$ contribute most to the quantity $\bar{a\bar B}$ depends on 
%1)
 the kernel,  $k_l$, and
  %2)
   $\ksm/\klg$. We explain these dependencies in turn.
The dependence on the kernel can be seen from the following.
We will  express $\bar{a\bar B}$ in $k$-space in terms of $\uGl$, $\utilde A$ and $\utilde B$.
First we focus on the Fourier transform of  ${a\bar B}$, which is given by the convolution
%\remark{EGB: edited above to reflect that equation below is FT of ${a\bar B}$ rather than $\bar{a\bar B}$, OK?}
\beq
(\utilde{a}*\bar{\utilde{B}})(k)
=\int dk'\ \utilde a(k')\utilde {\bar B}(k-k')
=\int dk'\ [1-\uGl(k')]\uGl(k-k') \utilde A(k')\utilde B(k-k').
\label{eqn_appx1}
\eeq
For fixed $k$, we can calculate which wavenumber $k'$ in the convolution contributes most by differentiating the factor $[1-\uGl(k')]\uGl(k-k')$ with respect to $k'$ and setting it to zero.
The solution $k'_0(k)$ depends on the form of the kernel $\uGl$.
How $k'_0(k)$ behaves at small $k$ is  of interest  because  we ultimately need to multiply $(\utilde{a*\bar B})(k)$ by $\uGl(k)$ to get the Fourier transform of  $\bar{a\bar B}$.
If $k'_0$  is small  for  small $k$, then we need only  consider the low wavenumber parts of $a$ and $\bar B$ because both $k'$ and $k-k'$ would be small in the integrand.
But $k'_0$ could in general be comparable to $k_l$ or even larger for small $k$.
For example, figure \ref{fig_k0} shows $k'_0(k)$ for a Gaussian kernel $\uGl(k)=e^{-k^2/2k_l^2}$ with $k_l/\klg=1$. For $k\leq k_l$, we see that $k'_0(k)$ is not small, and is of order  $k_l$.

%Although  $k'_0(k)$ measures where the maximum of the factor $[1-\uGl(k')]\uGl(k-k')$ lies,
However, if the spectra
%\remark{HZ changed below}
$\utilde A(k')$ or $\utilde B(k-k')$ vanishes near  $k'=k'_0(k)$ then the 
maximum contribution to equation (\ref{eqn_appx1}) 
must come from other wave numbers where $\utilde A$ and $\utilde B$ are non-vanishing.
% such as 
In the case of double peaked spectra, 
with peaks at $\klg$ and $\ksm$, 
the scale separation plays an important role in determining the significantly contributing wave numbers.
In the aforementioned example of figure \ref{fig_k0}, it is possible that  $\utilde A(k')\utilde B(k-k')$  in the integrand of equation (\ref{eqn_appx1}) vanishes at $k'=k'_0(k)\simeq k_l$ for small $k$.  That is, although $[1-\uGl(k')]\uGl(k-k')$ reaches its maximum at $k'_0(k)$ for small $k$,
%\remark{HZ deleted $\simeq \utilde A(k_l) \utilde B(-k_l)$}
$\utilde A(k')\utilde B(k-k')\simeq0$ there because of large scale separation.
 Indeed,  equation (\ref{fml2'}) is appropriate for cases with large scale separations between peaks, because the factor $[1-\uGl(k')]\uGl(k-k')$ cannot be  large at small $k$ and large $k'$.
Given a fixed small $k$, this factor  will vanish toward large $k'$ and retain some non-zero value at intermediate ($\simeq k_l$) and small ($\lesssim k_l$) $k'$ depending on the kernel.
Provided there is a large enough scale separation, the intermediate $k'$ regime  does not contribute since  $\utilde A$ and $\utilde B$ vanish there, leaving only  the small $k'$ part.

\begin{figure}
	\includegraphics[width=\columnwidth]{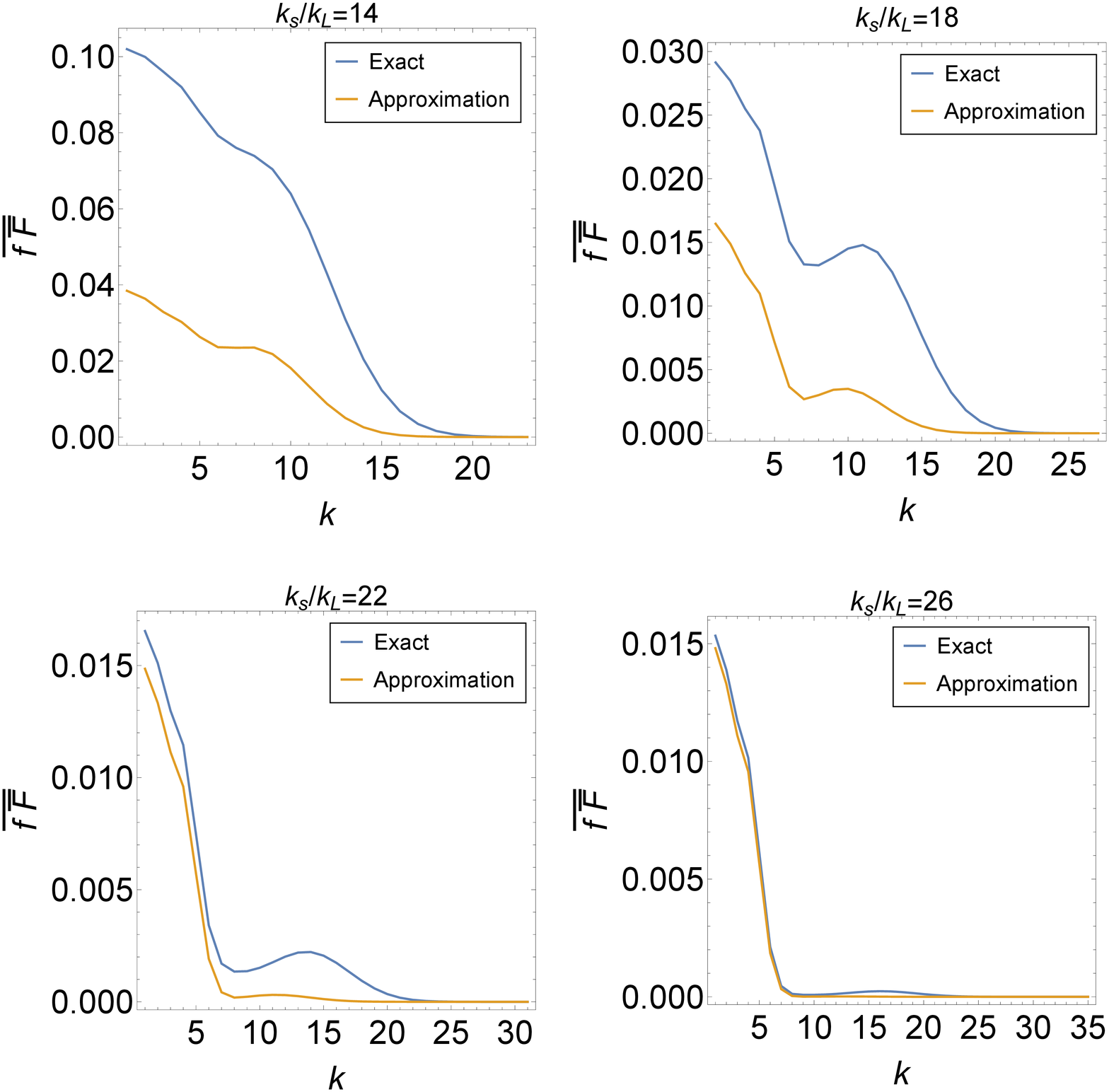}
		\caption{
		Comparisons of exact and approximated results of $\bar{f\bar{F}}$ for different scale separations $\ksm/\klg$.
	}
	\label{fig_test2.14a}
\end{figure}

\begin{figure}
	\centerline{
	\includegraphics[width=0.5\columnwidth]{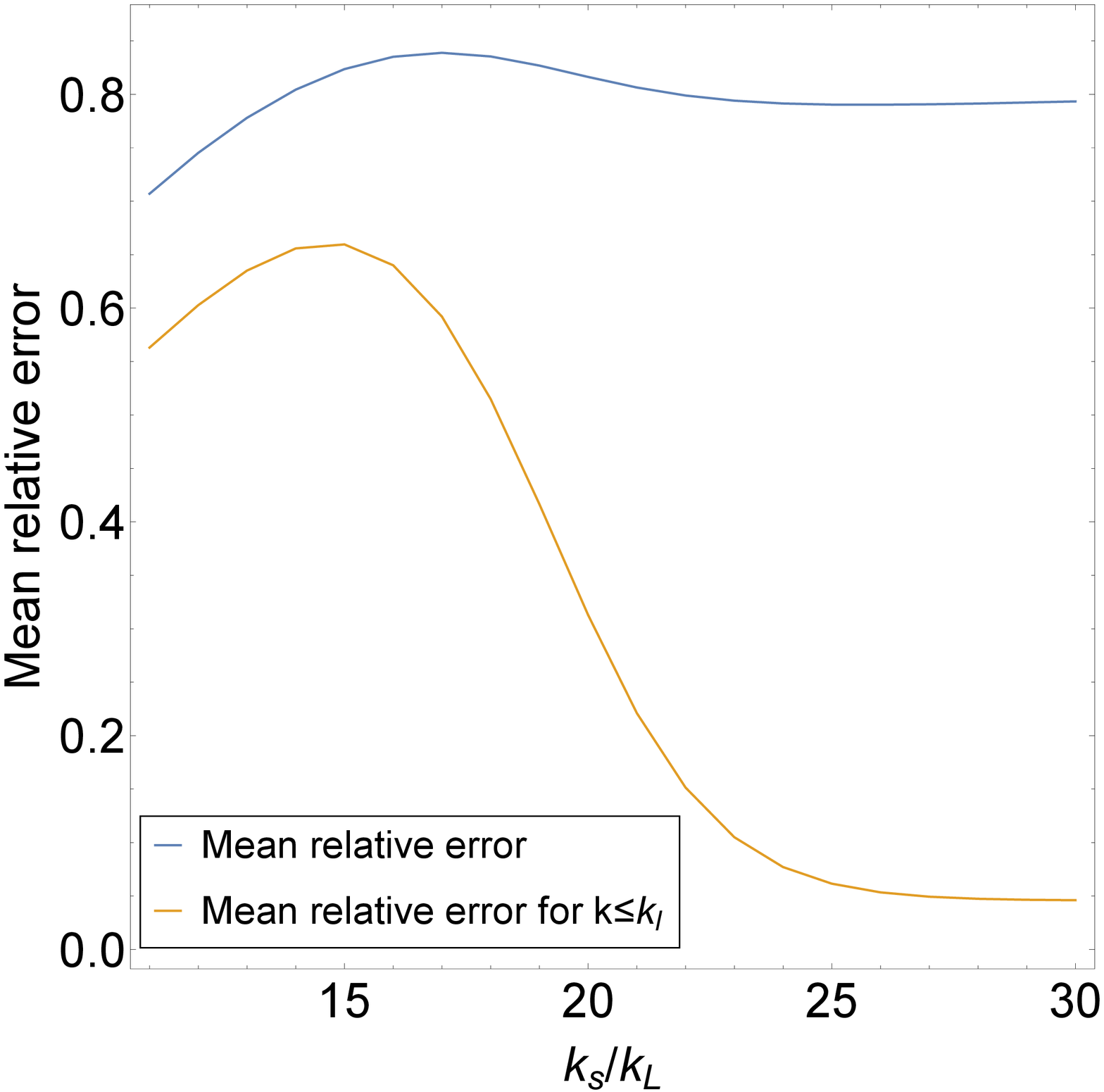}
	\includegraphics[width=0.5\columnwidth]{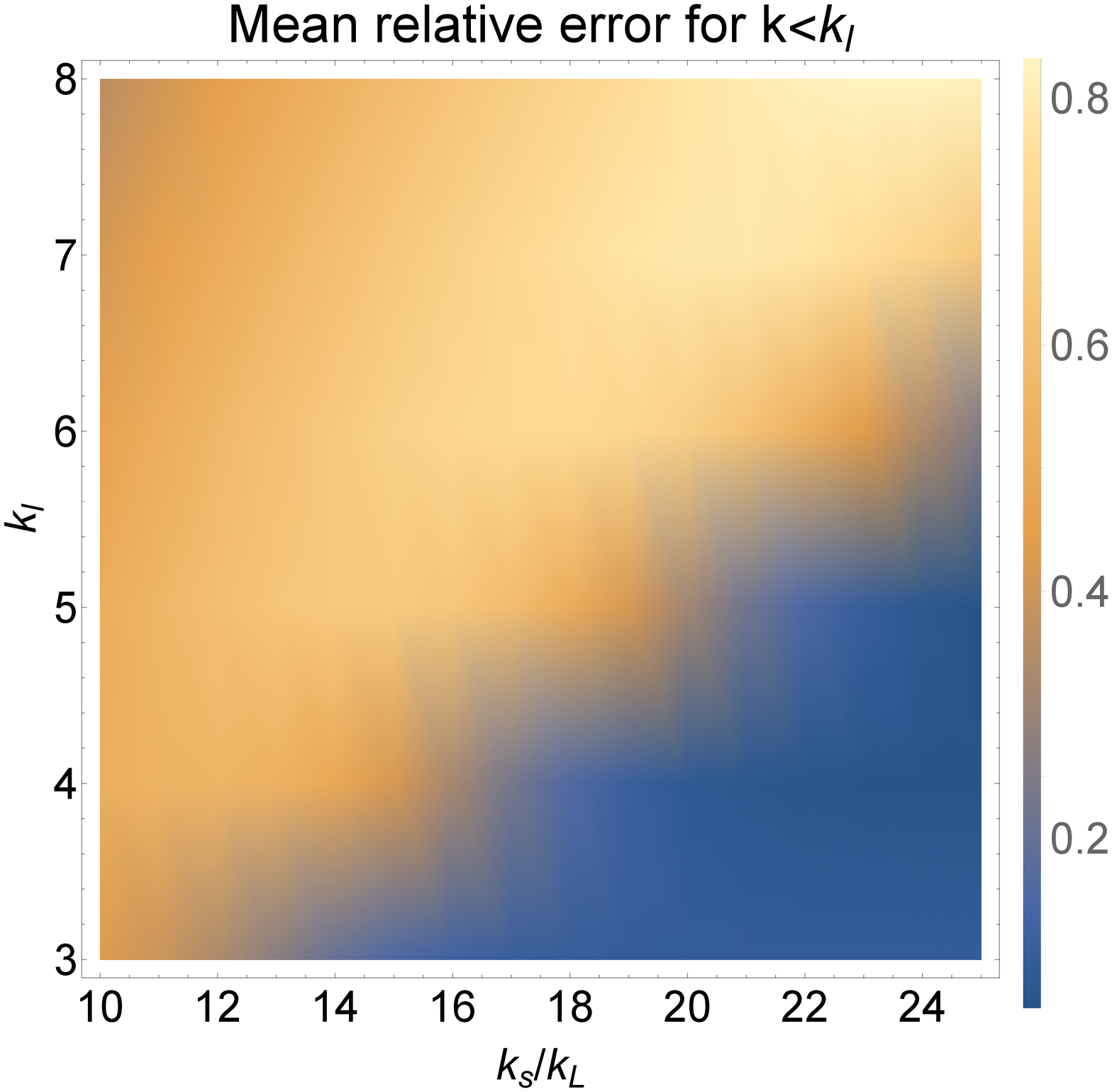}
}
	\caption{
		Left panel: The mean relative errors defined in equations (\ref{eqn:A4}) and (\ref{eqn:A5}) from comparing the exact and approximated results of $\bar{f\bar{F}}$ as a function of $\ksm/\klg$.
		Right panel: The mean relative error for $k<k_l$ as a function of $\ksm/\klg$ and $k_l$.
	}
	\label{fig_test2.14b}
\end{figure}

We quantify the importance of scale separation for the validity of equation (\ref{fml2'})
in figures \ref{fig_test2.14a} and \ref{fig_test2.14b} using the following double-peaked spectrum:
\beq
\utilde F(k)=\frac{1}{\sqrt{2\pi}\sigL}e^{-(k-\klg)^2/2\sigL^2}
+\frac{q}{\sqrt{2\pi}\sigs}e^{-(k-\ksm)^2/2\sigs^2},\ k\geq0;\ 
\utilde F(k)=\utilde F(-k),\ k<0
\eeq
where $\klg=1$, $\sigL=1$,  $\sigs=4$ and $q=4$ are fixed.
We use a Gaussian kernel for filtering, namely
\beq
\uGl(k)=e^{-k^2/2/k_l^2},
\eeq
where $k_l=5$ is fixed.
We then test equation (\ref{fml2'}) by comparing the exact result $\Pe=\mathcal{F}[{\bar{f\bar F}}]=G(k)[\utilde f*\utilde{\bar F}](k)$ and its approximation $\Pa=G(k)[\utilde{\bar F}*\utilde\gamma \utilde{\bar F}](k)$ [$\utilde\gamma$ is defined through equation (\ref{gamma2.4})] for different scale separations of the peaks, as quantified by $\ksm/\klg$.
The comparison  is  shown in figure \ref{fig_test2.14a} where blue curves are the exact results, and yellow ones are the approximations.

The efficacy of the approximation  can be quantified by the mean relative  difference between blue and yellow curves in the plots  figure \ref{fig_test2.14a}, that is
\beq
\bar\Delta=\frac
{\int_0^{k_\nu} dk\ \frac{\Pe-\Pa}{\Pe}}
{\int_0^{k_\nu} dk},
\label{eqn:A4}
\eeq
where we set $k_\nu=\ksm+2\sigs$.
The quantity $\bar\Delta$  as function of $k_s/k_L$ is  shown in blue in the left panel of figure \ref{fig_test2.14b}.
It remains relatively constant over the plot, even when scale separation is large. In that case, 
even though  the approximation agrees with the exact result at small $k$, 
%where the spectrum has significant values, 
 the relative deviation from the approximation becomes large  at large $k$.
But since we are interested in the net value of the  convolution at small $k\leq k_l$, 
 a better indicator of the efficacy of the approximation is the mean relative difference at $k\leq k_l$; that is
\beq
{\bar\Delta}_{k\leq k_l}=\frac
{\int_0^{k_l} dk\ \frac{\Pe-\Pa}{\Pe}}
{\int_0^{k_l} dk}.
\label{eqn:A5}
\eeq
This is shown in the yellow curve in the left panel of figure \ref{fig_test2.14b}.
$\Pa$ becomes a good approximation of $\Pe$ when $k_l=5$ and $\ksm/\klg\gtrsim 20$ (noting that $\klg=1$).
In the right panel of figure \ref{fig_test2.14b} we plot ${\bar\Delta}_{k\leq k_l}$ but now also varying $k_l$ in addition to $\ksm/\klg$.
The scale separation required to validate equation (\ref{fml2'}) increases with increasing $k_l$.

Note that the dependencies of correction terms to the mean field equations of Section \ref{sec_dynamo}  and the efficacy of the approximation (\ref{fml2'}) on scales are different:
The former  depends on the ratio $\klg/k_l$, whereas the latter depends on $\klg/\ksm$.
In the case of large scale separation, it is therefore possible that the error of the approximations are negligible but the MFE correction terms are still  significant.

\section{Derivation of the first two terms in equation (\ref{EMF2})}
\label{appx2}
The expansion rule (\ref{fml3}) cannot be immediately applied to the first term on the RHS of equation (\ref{eqn_dtub}) because $\bmb$ does not commute with the projection operator $\hat{\bf P}$.
%The key is to restore the distribution law of $\nabla^{-2}$ by noticing that we only need terms linear in $\bar\bmB$ and $\del\bar\bmB$.
Therefore let us write it as
\begin{align}
\epsilon_{ijk}\bar{
	b_k \hat P_{lj}
	(\bar B_n \partial_n b_l+b_n \partial_n \bar B_l)
}=
&\epsilon_{ijk}\bar{
	b_k \left(\delta_{lj}-{\partial_l\partial_j}{\nabla^{-2}}\right)
	(\bar B_n \partial_n b_l+b_n \partial_n \bar B_l)
}\notag\\
=&\epsilon_{ijk}\bar{
	b_k \left[
	(\bar B_n \partial_n b_j + b_n \partial_n \bar B_j)
	-{\partial_j}{\nabla^{-2}}(\partial_l \bar B_n \partial_n b_l + \partial_l b_n \partial_n \bar B_l)
	\right]
}\notag\\
=&\epsilon_{ijk}\bar{\bar B_n b_k\partial_n b_j}
+\epsilon_{ijk}\bar{b_k b_n \partial_n \bar B_j}
-2\epsilon_{ijk}\bar{b_k {\partial_j}{\nabla^{-2}}(\partial_l \bar B_n \partial_n b_l)}.
\label{eqn_B1}
\end{align}

The first term can be readily calculated assuming isotropy for turbulent fields, yielding
\beq
(1-\hg)\left(\frac{1}{3}\bar{\bmb\cdot\del\times\bmb}\ \bar B_i\right)
+\bar B_i \hg \left(\frac{1}{3}\bar{\bmb\cdot\del\times\bmb}\right).
\eeq

Denote the third term in equation (\ref{eqn_B1}) by $-2X_i$.
Then
\beq
X_i=\epsilon_{ijk}\bar{b_k {\partial_j}{\nabla^{-2}}(\partial_l \bar B_n \partial_n b_l)}
=\epsilon_{ijk}\bar{
	b_k {\nabla^{-2}} 
	(\partial_{jl}\bar B_n \partial_n b_l
	+\partial_l \bar B_n \partial_{jn} b_l)
}.
\eeq
The first term in the parentheses is $\ksm/\klg$ times smaller than the second, and is therefore dropped.
In Fourier space, the inverse of the Laplacian operator acting on the second term yields
\beq
\mathcal{F}[{\nabla^{-2}}(\partial_l \bar B_n \partial_{jn} b_l)]
=\frac{1}{k^2}
\int d^3 k'\ 
i(k_l-k'_l) \bar B_n(\bmk-\bmk')
(k'_j k'_n)b_l(\bmk').
\eeq
$\bmk-\bmk'$ is close to zero because of the presence of $\bar\bmB(\bmk-\bmk')$.
Therefore we expand $1/k^2$ as
\beq
\frac{1}{k^2}=\frac{1}{k'^2}+\mathO(|\bmk-\bmk'|).
\eeq
Only the zeroth order term is kept, because terms of higher order yield derivatives of $\bar\bmB$, which makes $X_i$ contain second or higher order derivatives of $\bar\bmB$.
Equivalently, this means the $\nabla^{-2}$ operator will not act on the $\bar\bmB$ term to this order.
%\remark{EB added "to this order"}
We now have, up to terms linear in $\bar\bmB$ or $\del\bar\bmB$,
\beq
X_i\simeq\epsilon_{ijk}\bar{\partial_l \bar B_n b_k {\partial_{jn}}{\nabla^{-2}}b_l}
\label{eqn_BX}
\eeq
using equation (\ref{fml3}).

Now the sum of the last two terms in equation (\ref{eqn_B1}) can be written as
\beq
\epsilon_{ijk}\bar{b_k b_n \partial_n \bar B_j}
-2\epsilon_{ijk}\bar{\partial_l \bar B_n b_k {\partial_{jn}}{\nabla^{-2}}b_l}
=(1-\hg)\left(\partial_l \bar{B}_n\xi_{iln}\right)
+\partial_l \bar{B}_n\hg\xi_{iln}
\label{eqn:B7}
\eeq
where
\beq
\xi_{iln}=\epsilon_{ijk}\bar{b_k\left(\delta_{jn}-2{\partial_{jn}}{\nabla^{-2}}\right)b_l}.
\eeq

To evaluate $\xi_{iln}$, note that its Fourier transform is proportional to
\beq
\epsilon_{ijk}\int d^3k'\ P_{kl}(k')\left(\delta_{jn}-2\frac{k'_j k'_n}{k'^2}\right)
\label{eqn:B9}
\eeq
since the helical part of $\bar{\utilde b_k\utilde b_l}$ ($\propto\epsilon_{pkl}k_p$) does not contribute.
Equation (\ref{eqn:B9}) then gives
\beq
\epsilon_{ijk}\int d^3k'\ \left(\delta_{kl}\delta_{jn}-2\delta_{kl}\frac{k'_j k'_n}{k'^2}-\delta_{jn}\frac{k'_k k'_l}{k'^2}\right)=0
\eeq
using $\int d\Omega' k_i k_j/k^2=\delta_{ij}/3$.
Therefore the RHS of equation (\ref{eqn:B7}) is zero and altogether we have
\beq
\epsilon_{ijk}\bar{
	b_k \hat P_{lj}
	(\bar B_n \partial_n b_l+b_n \partial_n \bar B_l)
}=
(1-\hg)\left(\frac{1}{3}\bar{\bmb\cdot\del\times\bmb}\ \bar B_i\right)
+\bar B_i \hg \left(\frac{1}{3}\bar{\bmb\cdot\del\times\bmb}\right).
\eeq

%=========================================================

\bibliographystyle{jpp}
\bibliography{meanfieldbib}%,RNGDNotes,blackmanjpp3-17-15bib}
%EB20 it seems that adding multiple files works and puts needed refs into .bbl with name of first file listed --never tried this before but just tried it. might make life easier 
\end{document}